\documentclass[a4paper,twocolumn,11pt,accepted=2020-04-20]{quantumarticle}
\pdfoutput=1
\usepackage[utf8]{inputenc}
\usepackage[english]{babel}
\usepackage[T1]{fontenc}
\usepackage{placeins}
\usepackage{afterpage}
\usepackage{hyperref}
\usepackage[numbers,sort&compress]{natbib}
\usepackage{url}

\usepackage{algpseudocode}
\usepackage[ruled]{algorithm}
\usepackage{mathtools}
\usepackage{standalone}
\usepackage{amssymb,amsmath,amstext}
\usepackage{graphicx}
\usepackage{epstopdf}
\usepackage{color}
\usepackage{appendix}
\usepackage[T1]{fontenc}
\usepackage{bbold}
\usepackage{bbm}
\usepackage{float}
\usepackage{latexsym}

\usepackage{hhline}
\renewcommand{\vec}[1]{\boldsymbol{#1}}  

\long\def\ca#1\cb{} 

\newcommand{\fattheta}{\boldsymbol{\theta}}
\newcommand{\tot}{\text{tot}}

\newcommand{\Tr}{{\rm Tr}}

\renewcommand{\leq}{\leqslant}

\newcommand{\noise}{\text{noise}}
\newcommand{\Var}{\text{Var}}

\DeclareMathOperator*{\argmax}{arg\,max}

\renewcommand{\vec}[1]{\boldsymbol{#1}}  

\makeatletter
\newcommand{\justified}{%
  \rightskip=10pt $\,$%
  \leftskip=10pt }
\makeatother

\begin{document}

\title{An Adaptive Optimizer for Measurement-Frugal Variational Algorithms}


\author{Jonas M. K{\"u}bler}
\affiliation{Theoretical Division, MS B213, Los Alamos National Laboratory, Los Alamos, NM 87545, USA.}
\affiliation{Max Planck Institute for Intelligent Systems, Max-Planck-Ring 4, 72076 T{\"u}bingen, Germany.}
\orcid{0000-0002-8634-7990}

\author{Andrew Arrasmith}
\affiliation{Theoretical Division, MS B213, Los Alamos National Laboratory, Los Alamos, NM 87545, USA.}
\orcid{0000-0003-2674-9370}

\author{Lukasz Cincio} 
\affiliation{Theoretical Division, MS B213, Los Alamos National Laboratory, Los Alamos, NM 87545, USA.}
\orcid{0000-0002-6758-4376}

\author{Patrick J. Coles} 
\email{pcoles@lanl.gov}
\affiliation{Theoretical Division, MS B213, Los Alamos National Laboratory, Los Alamos, NM 87545, USA.}
\orcid{0000-0001-9879-8425}

\begin{abstract}
Variational hybrid quantum-classical algorithms (VHQCAs) have the potential to be useful in the era of near-term quantum computing. However, recently there has been concern regarding the number of measurements needed for convergence of VHQCAs. Here, we address this concern by investigating the classical optimizer in VHQCAs. We introduce a novel optimizer called individual Coupled Adaptive Number of Shots (iCANS). This adaptive optimizer frugally selects the number of measurements (i.e., number of shots) both for a given iteration and for a given partial derivative in a stochastic gradient descent. We numerically simulate the performance of iCANS for the variational quantum eigensolver and for variational quantum compiling, with and without noise. In all cases, and especially in the noisy case, iCANS tends to out-perform state-of-the-art optimizers for VHQCAs. We therefore believe this adaptive optimizer will be useful for realistic VHQCA implementations, where the number of measurements is limited.
\end{abstract}
\maketitle

\section{Introduction}



There are various strategies to make use of noisy intermediate-scale quantum (NISQ) computers~\cite{preskill2018quantum}. One particularly promising strategy is to push most of the algorithmic complexity onto a classical computer while running only a small portion of the computation on the NISQ device. This is the idea behind variational hybrid quantum-classical algorithms (VHQCAs)~\cite{mcclean2016theory}. VHQCAs employ a quantum computer to efficiently estimate a cost function that depends on the parameters of a quantum gate sequence, and then leverage a classical optimizer to minimize this cost. VHQCAs intend to achieve a quantum advantage with NISQ computers by finding short-depth quantum circuits that at least approximately solve some problem. VHQCAs have been proposed for many applications including ground-state preparation, optimization, data compression, simulation, compiling, factoring, diagonalization, and others~\cite{peruzzo2014VQE, farhi2014QAOA, johnson2017qvector, romero2017quantum, larose2018, arrasmith2019variational, cerezo2019variational, jones2019variational, yuan2018theory, li2017efficient, kokail2019self, Khatri2019quantumassisted, jones2018quantum, heya2018variational, endo2018variational,sharma2019noise, carolan2019variational,yoshioka2019variational,bravo-prieto2019,xu2019variational,mcardle2019variational,cirstoiu2019variational}.


A concern about VHQCAs is that they might require prohibitively many quantum measurements (shots) in order to achieve convergence of the cost function~\cite{troyer2015}, especially for applications like quantum chemistry that require chemical accuracy~\cite{cao2018quantum, mcardle2018quantum}. In response to this concern, there has been an recent explosion of papers looking to improve the measurement frugality of VHQCAs by simultaneously measuring commuting subsets of the Pauli operators needed for the cost function~\cite{Jena2019,Izmaylov2019,Yen2019,Gokhale2019,Crawford2019,Gokhale2019-2, huggins2019efficient}. 


Here, we approach the problem from a different direction by aiming to improve the classical optimizer. There have been several recent efforts to improve optimizers for VHQCAs~\cite{verdon2019learning, wilson2019optimizing, nakanishi2019, parrish2019, stokes2019quantum}. Our approach is different from these works in that the optimizer we propose is specifically constructed to achieve measurement frugality. In particular, we develop an adaptive optimizer that is adaptive in two senses: it frugally adjusts the number of shots for a given iteration and for a given partial derivative. 

Our method is inspired by the classical machine learning algorithm named Coupled Adaptive Batch Size (CABS)~\cite{Balles2017}. For pedagogical reasons, we first directly adapt the CABS algorithm to VHQCA applications and call the resulting algorithm Coupled Adaptive Number of Shots (CANS). In order to achieve greater measurement frugality, we go beyond direct adaptation and modify the optimizer to account for differences in the number of shots needed to estimate individual components of the gradient. We call this method individual-CANS (iCANS).



While iCANS is conceptually simple, it nevertheless performs very well. Using IBM's simulator~\cite{gadi_aleksandrowicz_2019_2562111}, we implement iCANS and other state-of-the-art optimizers such as Adam~\cite{Kingma2015}, SPSA~\cite{spall1992}, and sequential gate optimization~\cite{parrish2019, nakanishi2019} for both the variational quantum eigensolver~\cite{peruzzo2014VQE} and variational quantum compiling~\cite{Khatri2019quantumassisted, jones2018quantum, heya2018variational, sharma2019noise}.  We find that iCANS on average performs the best. This is especially true for our implementations in the presence of noise, i.e., with IBM's simulator of their NISQ device. This is encouraging since VHQCAs must be able to run in the presence of noise to be practically useful.

Ultimately, one can take a multi-pronged approach to reducing measurements in VHQCAs, e.g., by combining our measurement-frugal classical optimizer with the recent advances on Pauli operator sets in Refs.~\cite{Jena2019,Izmaylov2019,Yen2019,Gokhale2019,Crawford2019,Gokhale2019-2, huggins2019efficient}. However, one can apply our optimizer to VHQCAs that do not involve the measurement of Pauli operator sets (e.g., the VHQCAs in \cite{larose2018, arrasmith2019variational, cerezo2019variational}). In this sense, our work is relevant to all VHQCAs.

In what follows, we first give a detailed review of various optimizers used in the classical machine learning and quantum circuit learning literature. We remark that this lengthy review aims to assist readers who may not have a background in classical optimization, as this article is intended for a quantum-computing audience. (Experienced readers can skip to Section~\ref{sec:algorithm}.) We then present our adaptive optimizer, followed by the results of our numerical implementations.

\section{Background}

\subsection{Gradient Descent}

One standard approach to minimization problems is gradient descent, where the optimizer iteratively steps along the direction in parameter space that is locally ``downhill'' (i.e., decreasing) for some function $f(\fattheta)$. Mathematically, we can phrase the step at the $t$-th iteration as
\begin{align}
    \fattheta^{(t+1)} = \fattheta^{(t)} - \alpha \vec{\nabla} f(\fattheta^{(t)}),
\end{align}
where $\alpha$ is called the \textit{learning rate}. If one takes a large learning rate, one cannot be sure that one will not go too far and possibly end up at a higher point. For a small learning rate one is more guaranteed to keep making incremental progress (assuming the change in slope is bounded), but it will take much longer to get to a minimum. Knowing an upper bound on the slope is therefore very helpful in determining the appropriate learning rate. 

To formalize this discussion, we review the notion of Lipschitz continuous gradients. The gradient of a function $f$ is Lipschitz continuous if there exists some $L$ (called the Lipschitz constant) such that 
\begin{align}\label{eq:Lipschitz}
    \|\vec{\nabla} f(\fattheta^{(t+1)} - \vec{\nabla} f(\fattheta^{(t)})  \| \leq L \|\fattheta^{(t+1)} -\fattheta^{(t)}\|,
\end{align}
for all $\fattheta^{(t+1)}$ and $\fattheta^{(t)}$. (We note that in our notation the $\| \cdot\|$ without a subscript denotes the $\ell_2$ or Euclidean norm.) When this holds, we can see that the fractional change in the gradient over the course of one step is bounded by $\alpha L$, meaning that for sufficiently small $\alpha$ we can be sure that we are following the gradient. In fact, the convergence of the basic gradient descent method is guaranteed for deterministic gradient evaluations so long as $\alpha<2/L$~\cite{Balles2017}. In machine learning contexts $L$ is usually unknown, but for VHQCAs it is often possible to determine a good bound. We discuss this alongside an analytic formula for estimating gradients for VHQCAs in the next subsection.

\subsection{Gradient Estimation}

Working with the exact gradient is often difficult for two reasons. First the gradient can depend on quantities that are expensive to estimate with high precision. Second, it might be that no analytic form for the gradient formula is accessible, and hence the gradient must be approximated by finite differences. In the following we discuss the two scenarios in more detail.

\subsubsection{Analytic gradients}

If one has sufficient knowledge of the structure of the optimization problem under consideration, it might be possible to find analytic expressions for the gradient of the function. In deep learning this is what is provided via the backpropagation algorithm, which allows one to take analytic derivatives with respect to all parameters \cite{lecun2015}. However these formulas are usually expressed as an average over the full sample one has available in a learning task. To decrease the cost of evaluating the gradient often only a subset of the full sample, a so-called mini-batch, is used to get an unbiased estimate of the gradient \cite{lecun2015}. This introduces a trade-off between the cost of the gradient estimation and its achieved precision.

In VHQCAs there exist similar scenarios where it is possible to analytically compute the gradients \cite{Mitarai2018, Schuld2019,bergholm2018pennylane}. For example if the parameters describe rotation angles of single-qubit rotations and the cost function is the expectation value of some operator $\boldsymbol{A}$, $f=\left<\boldsymbol{A}\right>$, partial derivatives can be computed as 
\begin{equation}\label{eq:analytic_gradient}
    \partial_{\theta_i} f(\boldsymbol{\theta}) =\frac{ f(\boldsymbol{\theta}+\frac{\pi}{2}\hat{e}_i) -f(\boldsymbol{\theta}-\frac{\pi}{2}\hat{e}_i)}{2},
\end{equation}
i.e., the partial derivative is determined by the value of the cost function if one changes the $i$-th component by $\pm \pi/2$. However, the value of the cost function can only be estimated from a finite number of measurements, and this number of measurements as well as the noise level of the computation itself determine the precision of the gradient estimates. Therefore it is important to understand how to choose the number of shots, and keep in mind that for VHQCAs the gradient estimate is always noisy to some extent, even though it is referred to as analytical.

An immediate extension of this is that \eqref{eq:analytic_gradient} can be used recursively to define higher derivatives. This result then allows one to determine a usefully small upper bound on $L$ in \eqref{eq:Lipschitz}. In particular, we note for operators with bounded eigenspectra, the largest magnitude of a derivative of any order we can find with \eqref{eq:analytic_gradient} is precisely half the difference between the largest and smallest eigenvalues $\lambda_{\max}$ and $\lambda_{\min}$, respectively. Thus,
\begin{equation}\label{eq:tighter_lipschitz}
    L \leq \frac{\lambda_{\max}-\lambda_{\min}}{2}.
\end{equation}
For the common case where the eigenspectrum is unknown but we know how to decompose $\boldsymbol{A}$ into a weighted sum over tensor products of Pauli matrices, $\boldsymbol{A}= \sum_i a_i \boldsymbol{\sigma}_i $, we can bound the highest and lowest eigenvalues in turn by $\sum_i| a_i |$ and $-\sum_i| a_i |$, respectively, which gives
\begin{equation}\label{eq:general_lipschitz}
    L \leq \sum_i| a_i |.
\end{equation}
By setting equality in \eqref{eq:general_lipschitz} (or \eqref{eq:tighter_lipschitz} when we have more information), we therefore find a useful Lipschitz constant.

\subsubsection{Finite Differencing}

If one does not have access to analytical gradients, one way to approximate the partial derivatives is by taking a finite $\delta$ step in parameter space
\begin{equation}\label{eq:finite_difference_gradient}
    \partial_{\theta_i} f(\boldsymbol{\theta}) \approx \frac{ f(\boldsymbol{\theta}+\delta\hat{e}_i) -f(\boldsymbol{\theta}-\delta\hat{e}_i)}{2\delta}.
\end{equation}
Again, as in the analytical case, the function values need to be estimated by a finite number of shots introducing statistical noise. However, as opposed to the analytic case, the estimate \eqref{eq:finite_difference_gradient} is systematically wrong, with an error that scales with $\delta^2$. Therefore, one might want to decrease the parameter $\delta$ during an optimization procedure using such a gradient estimate. Intuitively this makes the optimization harder, and was recently discussed in the context of VHQCAs~\cite{harrow2019}.

\subsection{Noisy Gradient Descent}

For the case where one has noise in one's measurement of the gradient, the analysis of a gradient descent procedure becomes more complicated as the best one can achieve are statements about the behavior that can be expected on average. However, so long as one's estimates are unbiased (i.e., repeated estimates average to the true gradient) one should still end up near a minimum. This idea is at the heart of all stochastic gradient descent methods which we discuss now.

\subsubsection{Stochastic/Mini-Batch Gradient Descent}
In cases such as VHQCAs (as well as some machine learning applications), we cannot access the gradients directly and therefore need to estimate the the gradients by sampling from some distribution. A standard approach to this case is to choose some number of samples that are needed to achieve a desired precision. This method is known as either stochastic or mini-batch gradient descent. (A mini-batch here refers to a collection of samples, usually much smaller than the total population.)

The number of samples as well as the learning rate are usually set heuristically, in order to balance competing interests of efficiency and precision. First, when collecting samples is computationally expensive, it can sometimes be more efficient to take less accurate gradient estimates in order to converge faster, though doing so can be detrimental if it means that one ends up needing to perform an inordinate number iterations~\cite{Guerreschi2017}. Second, it does not make sense to attempt to achieve a precision greater than intrinsic accuracy of the distribution from which one samples. If there is some error expected in the representation of the distribution one samples the gradients from, there is therefore an upper bound on the number of samples that it is sensible to take based on that accuracy~\cite{Guerreschi2017}. For the case of VHQCAs, this often means that the upper limit on the number of samples, $s_{\max}$ depends on the (usually unknown) bias $b_{\noise}$ introduced to the gradient measurements by the noise of the physical quantum device:
\begin{equation}
    s_{\max}\simeq \frac{\Var(f(\fattheta))}{b_{\noise}^2(\fattheta))}.
\end{equation}
Since for VHQCAs this bias is a function of the unknown, time varying device noise for the specific gate sequence, often the best one can do is to make a rough estimate about its order of magnitude and use that in the denominator.

Typically, the number of samples as well as the learning rate are heuristically adjusted based on the structure of the cost landscape as well as the error level. When little information is known about the optimization problem, the minimization process is optimized either by manual trial and error until an acceptable choice is found or using a hyper-parameter optimization strategy~\cite{Bergstra2011}.

For a stochastic gradient approach to converge quickly, it is often helpful to decrease the error in the optimization steps during the run of the optimization. This can be done by either decreasing the learning rate $\alpha$, or minimizing the noise in the gradient estimates. The following two subsections introduce two methods from machine learning that respectively take these two strategies.

\subsubsection{Adam}
Adam is a variant of stochastic gradient in which the step that is taken along each search direction is adapted based on the first and second moment of the gradient \cite{Kingma2015}. To do this, one takes an exponential decaying average of the first and second moment ($m_t$ and $v_t$, respectively) for each component of the gradient individually 
\begin{align}
    m_t &= \beta_1 m_{t-1} + (1-\beta_1) g_t\\
    v_t &= \beta_2 v_{t-1} + (1-\beta_2) g_t^2,
\end{align}
where the square is understood element-wise, $g_t$ is the gradient estimate at step $t$, and $\beta_1, \beta_2$ are the constants that determine how slowly the variables are updated. The parameters are then updated with the following rule:
\begin{align}
    \fattheta^{(t+1)} = \fattheta^{(t)} - \alpha \frac{\hat{m}_t}{\sqrt{\hat{v}_t} + \epsilon},
\end{align}
where $\hat{m}_t$ ($\hat{v}_t$) is an initialization-bias-corrected version of $m_t$ ($v_t$), and $\epsilon$ is a small constant to ensure stability \cite{Kingma2015}. One particular feature of Adam is that the adaptation happens individually for each component of the gradient. We also briefly mention that there is a recent modification to Adam that looks promising, called Rectified Adam (RAdam)~\cite{liu2019variance}. RAdam essentially selectively turns on the adaptive learning rate once the variance in the estimated gradient becomes small enough.

While Adam has made a large impact in deep learning, to our knowledge it has not been widely considered in the context of VHQCAs.

\subsubsection{CABS}

Balles et al.~analyzed the problem of choosing the sample size in the context of optimizing neural networks by stochastic gradient descent~\cite{Balles2017}. Their approach is to find the number of samples $s$ that maximizes the expected gain per sample at each iteration. 

In the following we denote the  $i$-th component of the estimated gradient by $g_i$, the empirical variance of the estimate $g_i$ by $S_i$, the actual gradient by $\vec{\nabla} f$, and the actual covariance matrix (in the limit of infinite samples or shots) of the gradient estimation by $\Sigma$. Balles et al.~introduce a lower bound $\mathcal{G}$ on the gain (improvement in the cost function) per iteration. Accounting for the finite sampling error, they find that the average value of $\mathcal{G}$ is~\cite{Balles2017}
\begin{equation}\label{eq:expected_gain}
    \mathbb{E}\left[\mathcal{G}\right] = \left( \alpha - \frac{L \alpha^2}{2}\right)\|\vec{\nabla} f\|^2- \frac{L\alpha^2}{2s}\Tr (\Sigma).
\end{equation}
As an immediate consequence, they then find that the expected gain at any step has a positive lower bound if
\begin{equation}\label{eq:alpha_bound}
    \alpha\leq \frac{2\|\vec{\nabla} f\|^2}{L\left(\|\vec{\nabla} f\|^2+\Tr (\Sigma)/s \right)}.
\end{equation}

By taking a small but fixed $\alpha$, Balles et al. then maximize the lower bound on the expected gain per sample by taking
\begin{align}\label{eq:optimal_shots_exact}
    s = \frac{2L\alpha}{2-L\alpha} \frac{\text{Tr}(\Sigma)}{\|\vec{\nabla} f\|^2}
\end{align}
samples~\cite{Balles2017}. Unfortunately, this formula depends on quantities $\Sigma$ and $\vec{\nabla} f$ that are not accessible. Therefore in CABS, $\Sigma$ is replaced by an estimator $\hat{\Sigma}$ and, specializing to the case where the minimum value of $f$ is known to be zero, $\|\vec{\nabla} f\|^2$ is replaced by $f/\alpha$ as the gradient estimator is biased. Since the Lipschitz constant is also often unknown in the machine learning problems they were considering, they also drop the factor of $2L\alpha/(2-L\alpha)$~\cite{Balles2017}. CABS then proceeds as a stochastic gradient descent with a fixed learning rate and a number of samples that is selected at each iteration based on \eqref{eq:optimal_shots_exact} with the quantities measured at the previous point, making the assumption that the new point will be similar to the old point. 

As discussed in the next section, our adaptive optimizer for VHQCAs is built upon the ideas behind CABS (particularly \eqref{eq:optimal_shots_exact}), although our approach differs somewhat.

\subsection{SPSA}
The simultaneous perturbation stochastic approximation (SPSA) algorithm \cite{spall1992} is explicitly designed for a setting with only noisy evaluation of the cost function, where no analytic formulas for the gradients are available. It is also a descent method, however, instead of estimating the full gradient, a random direction is picked and the slope in this direction is estimated. Based on this estimate a downhill step in the sampled direction is taken:
\begin{align}
    \fattheta^{(t+1)} = \fattheta^{(t)} - \alpha_t \boldsymbol{g}(\fattheta^{(t)}). 
\end{align}
Here $\boldsymbol{g}(\fattheta^{(t)})$ is the estimated slope in the random direction and estimated as \cite{spall1998}:
\begin{equation}\label{eq:SPSA_gradient}
    \boldsymbol{g}(\fattheta^{(t)}) = \frac{f(\fattheta^{(t)} + c_t \boldsymbol{\Delta}_t) - f(\fattheta^{(t)} - c_t \boldsymbol{\Delta}_t)}{2 c_t} \boldsymbol{\Delta}_t^{-1},
\end{equation}
where $\boldsymbol{\Delta}_t $ is the random direction sampled for the $t$-th step and $\boldsymbol{\Delta}_t^{-1}$ simply denotes the vector with its element-wise inverses. In order to ensure convergence the finite difference parameter $c_t$ as well as the learning rate $\alpha_t$ have to be decreased over the optimization run. This is commonly done by using a prefixed schedule \cite{spall1998}. In this approach, we have
\begin{equation}
     \alpha_t=\frac{\alpha_0}{(1+k)^\beta} \qquad\text{and} \qquad
     c_t = \frac{c_0}{(1+k)^\gamma}\,.
\end{equation}

In the original formulation, the idea is usually to estimate the cost function in \eqref{eq:SPSA_gradient} by a single measurement. However, in a quantum setting it seems intuitive to take a larger number of measurements for the estimation, as was done in \cite{kandala2017}.

\subsection{Sequential Subspace Search}
Another approach to optimizing a multivariate cost function is to break the problem into sub-parts which are independently easier to handle. The generic idea is to define a sequence of subspaces of parameter space to consider independently. These methods then approach a local minimum by iteratively optimizing the cost function on each subspace in the sequence. Now we discuss two instances of this approach: the famous Powell method~\cite{powell1964} as well as a recently proposed method specialized to VHCQAs~\cite{nakanishi2019, parrish2019}.

\subsubsection{Powell Algorithm}
The Powell algorithm~\cite{powell1964} is a very useful gradient-free optimizer that specializes the subspace search to the case of sequential line searches. Specifically, starting with some input set of search vectors $V=\{\vec{v}_i\}$ (often the coordinate basis vectors of the parameter space) this method sequentially finds the set of displacements $\{a_i\}$ along each search vector that minimizes the cost function. Next, the method finds the $\vec{v}_j$ associated with the greatest displacement, $a_j=\max({a_i})$. This $\vec{v}_j$ is then replaced with the total displacement vector for this iteration, namely:
\begin{equation}
    \vec{v}_j\to \sum_i a_i \vec{v}_i,
\end{equation}
and then the next iteration begins with this updated set of search vectors. This replacement scheme accelerates the convergence and prevents the optimizer from being trapped in a cyclic pattern.  In practice, the displacements $a_i$ are typically found using Brent's method~\cite{Brent2013}, but in principle any gradient-free scalar optimizer could work. (Gradient-based scalar optimizers would make Powell's method no longer ``gradient-free.'')

\subsubsection{Sequential Optimization by Function Fitting}

In the special case of VHQCAs where the cost function is expressed as an expectation value of some Hermitian operator and the quantum circuit is expressed as fixed two-qubit gates and variable single-qubit rotations, it is possible to determine the functional form of the cost function along a coordinate axis~\cite{nakanishi2019}. After fitting a few parameters, it becomes possible to compute where the analytic minimum should be in order to find the optimal displacement along any given search direction. This can be scaled up to finding the analytic minimum (exact up to distortions from noise) on some subspace that is the Cartesian product of coordinate axes, though this is hampered by the fact that the number of parameters that must be fit scales exponentially with the dimension of the subspace~\cite{nakanishi2019}. We will refer to this algorithm as the Sequential Optimization by Function Fitting (SOFF) algorithm. We note that a very similar method was published shortly after SOFF~\cite{parrish2019}. The primary difference was the incorporation of the Anderson and Pulay convergence acceleration procedures used in computational  chemistry~\cite{anderson1965iterative,pulay1982improved}.

We note that, though SOFF and Powell are closely related, due to the limitation to only searching along coordinate axes, it is not possible to take arbitrary search directions, thus SOFF is not quite a special case of Powell's method. For VHQCA problems where it is applicable, SOFF has been demonstrated to be highly competitive with or better than other standard optimization schemes like Powell's method~\cite{nakanishi2019,parrish2019}.

\section{Adaptive Shot Noise optimizer}\label{sec:algorithm}

As mentioned above, the basic idea behind our approach is similar to that of CABS~\cite{Balles2017}, but we implement those ideas in a different way. Specifically, by implementing different estimates for the inaccessible quantities in \eqref{eq:optimal_shots_exact} that are suitable to the number of shots in a VHQCA (rather than the batch size in a machine learning method), we arrive at a variant of CABS we name \textit{Coupled Adaptive Number of Shots} (CANS). Recognizing that a different number of shots might be optimal for estimating each component of the gradient in VHQCAs, we further develop this variation into \textit{individual-CANS} (iCANS), which is our main result. For pedagogical purposes, we first introduce CANS and then present iCANS.

\begin{figure}
\begin{algorithm}[H]
\begin{algorithmic}[1]
\Statex \textbf{Input:} Learning rate $\alpha$, starting point $\fattheta_0$, min number of shots per estimation $s_{\min}$, number of shots that can be used in total $N$, Lipschitz constant $L$, running average constant $\mu$, bias for gradient norm $b$
\State initialize: $\fattheta \gets \fattheta_0 $, $s_{\tot} \gets 0$,
$\vec{s} \gets (s_{\min} ,... ,s_{\min})^T$, $\vec{\chi}' \gets (0,...,0)^T$,
$\vec{\xi}' \gets (0,...,0)^T$, $k\gets 0$
\While{$s_{\tot} < N$}
    \State $\vec{g}, \vec{S} \gets iEvaluate(\fattheta, \vec{s})$
\State $s_{\tot} \gets s_{\tot} + 2 \sum_i s_i$
    \State $\vec{\xi}_\ell' \gets \mu \vec{\xi}_\ell' + (1-\mu) \vec{S}_\ell$
    \State $\vec{\chi}_\ell' \gets \mu \vec{\chi}_\ell' + (1-\mu) \vec{g}_\ell$ 
    \State $\vec{\xi}_\ell \gets \vec{\xi}_\ell'/(1-\mu^{k+1})$
    \State $\vec{\chi}_\ell \gets  \vec{\chi}_\ell'/(1-\mu^{k+1})$ 
\For{$ i \in [1,...,d]$}
    \If{iCANS1}
        \State $\fattheta_i \gets \fattheta_i - \alpha \vec{g}_i$
    \ElsIf{iCANS2}
        \If{$\alpha \leq \frac{g_i^2}{L\left(g_i^2+S_i/s_i + b \mu^k\right)}$}
            \State $\fattheta_i \gets \fattheta_i - \alpha \vec{g}_i$
        \Else
            \State $\alpha' \gets \frac{g_i^2}{L\left(g_i^2+S_i/s_i + b \mu^k\right)}$
            \State $\fattheta_i \gets \fattheta_i - \alpha' \vec{g}_i$
        \EndIf
    \EndIf
    \State $s_i \gets \left\lceil\frac{2L\alpha}{2-L\alpha} \frac{\xi_i}{\chi_i^2 + b \mu^k}\right\rceil$
    \State $\gamma_i \gets \frac{1}{s_i} \left[\left(\alpha - \frac{L\alpha^2}{2}\right) \chi_i^2 - \frac{L \alpha^2}{2 s_i} \xi_i\right]$
\EndFor
\State $s_{\max} \gets s_{\argmax_i \gamma_i}$ 
\State $\vec{s} \gets clip(\vec{s}, s_\text{min} , s_\text{max} )$ 
\State $k\gets k + 1$
\EndWhile
\end{algorithmic}
\caption{\justified{Stochastic gradient descent with iCANS1/2. The function {$iEvaluate(\fattheta, \vec{s})$} evaluates the gradient at $\fattheta$ using $s_i$ shots for the $i$-th derivative via the parameter shift rule \eqref{eq:analytic_gradient}. This function returns the estimated gradient vector $\vec{g}$ as well as the vector $\vec{S}$ whose components are the variances of the estimates of the partial derivatives.}}
\label{alg:iCANS}
\end{algorithm}
\end{figure}
\subsection{CANS}
We now discuss our adaptation of CABS to the setting of VHQCAs. In order to use the number of shots recommended by the CABS method, we need to rewrite \eqref{eq:optimal_shots_exact} using only quantities that are accessible. Making use of the parameter shift rule \eqref{eq:analytic_gradient}, we have access to the Lipschitz constant $L$ given by \eqref{eq:general_lipschitz}. An unbiased estimate of $\text{Tr}(\Sigma)$ is given by $\sum_{i=1}^d S_i = \|S\|_1$, i.e., by the empirical variances of the gradient components. (Here and below $d$ is the number of parameters being optimized.) The naive estimate of $\|\vec{\nabla} f\|^2$ is $\|g\|^2$, with $g := (g_1,..., g_l)^T$ the estimated gradient. This estimator is biased (see Equation (17) of \cite{Balles2017}), however using a bias-corrected version is numerically unstable.
With these choices, we then define CANS as the CABS algorithm with \eqref{eq:optimal_shots_exact} replaced by
\begin{align}\label{eq:CANS}
    s = \frac{2L\alpha}{2-L\alpha} \frac{\|S\|_1}{\|\vec{g}\|^2}.
\end{align}
We note that the learning rate $\alpha$ must be less than $2/L$ with this formalism. The CANS algorithm is included in Appendix~\ref{App:CANS} for completeness. For the remainder of the paper we will focus on iCANS, which we introduce next.

\subsection{iCANS}

The CANS algorithm is inspired by CABS \cite{Balles2017}, which was designed for applications in deep learning. Therein for each data point the full gradient is evaluated, and noise arises by considering only a minibatch of the full sample. In VHCQAs, however, each individual partial derivative is estimated independently. This gives us the freedom to distribute measurements over the estimation of the partial derivatives more effectively. This is the idea behind iCANS, which is shown in Algorithm~\ref{alg:iCANS} and described below.


iCANS prioritizes the individual partial derivatives rather than the gradient magnitude as in \eqref{eq:expected_gain}. For this purpose, we define $\mathcal{G}_i$ as our lower bound on the gain (i.e., the improvement in the cost function) associated with the change in parameter $\theta_i$ for a given optimization step. Furthermore, we define $\gamma_i$ as the expected gain per shot (i.e., the expectation value of $\mathcal{G}_i$ divided by the number of shots) as follows:
\begin{align}\label{eq:gain_per_shot}
   \gamma_i :=\frac{\mathbb{E}\left[\mathcal{G}_i\right]}{s_i} =\frac{1}{s_i} \left[\left(\alpha - \frac{L\alpha^2}{2}\right) g_i^2 - \frac{L \alpha^2}{2 s_i} S_i\right],
\end{align}
where $s_i$ is the suggested number of shots for the estimation of the $i$-th partial derivative. Note that \eqref{eq:gain_per_shot} is an adaptation of \eqref{eq:expected_gain} to our setting.

In analogy with the CANS approach (see \eqref{eq:CANS}), we estimate the number of shots that maximizes \eqref{eq:gain_per_shot} with
\begin{align}\label{eq:shots_iCANS}
    s_i = \frac{2L\alpha}{2-L\alpha} \frac{S_i}{g_i^2}.
\end{align}
As with CANS, we again note that this formalism is only valid if $\alpha<2/L$. The idea now is to update each parameter with a gradient-descent step, where each partial derivative is estimated with its individual optimal number of shots. However, empirically those parameters that are close to a local optimal value (hence have a small expected gain) require a large number of shots, while parameters that are far from convergence (and hence usually have a large expected gain) require a small number of shots. We therefore restrict our algorithm to not take more shots for any partial derivative than a cap we will call $s_{\max}$. We take $s_{\max}$ to be the number of shots needed in order to estimate the partial derivative for the parameter $\theta_{i_{\max}}$, where $i_{\max}$ is the index associated with highest expected gain per shot. In other words:
\begin{align}
    i_\text{max} &= \argmax_i \left(\gamma_i\right),\\
    s_\text{max} &= s_{i_\text{max}},
\end{align}
and we impose $s_i\leq s_{\max}$ for all partial derivatives.

\afterpage{\FloatBarrier}
We note that the introduction of this cap on the number of shots is a heuristic choice which we find to often be beneficial to shot frugality, but which removes the guarantee that $\gamma_i$ will be maximized or even positive. In order to preserve this frugality while retaining the guarantee of positive expected gains, one can also introduce a step that verifies that the learning rate to be used is appropriate after the measurements are taken and adapts it if it is not. Motivated by \eqref{eq:alpha_bound}, we check the following condition for each component of the gradient:
\begin{equation}
\label{eq:learn_rate_check}
    \alpha \leq \frac{g_i^2}{L\left(g_i^2+S_i/s_i \right)}.
\end{equation}
When this condition fails to hold for the $i$-th partial derivative, we temporarily replace $\alpha$ with the right hand side of \eqref{eq:learn_rate_check} for the update along that direction. Adding in this check results in a more conservative procedure as it takes smaller steps when needed in order to enforce that $\gamma_i > 0$, and thus restores the guarantee that $\mathbb{E}\left[\mathcal{G}\right]>0$. Below, we will refer to iCANS without this learning rate check as iCANS1 and with it as iCANS2. The distinction between iCANS1 and iCANS2 is made in Algorithm~\ref{alg:iCANS} with the conditional statements on lines 10 and 12.

Beyond the core components of iCANS given above, both implementations of iCANS also take two more hyperparameters for increased stability. Since iCANS is intended to be deployed on highly noisy problems, we find that it is beneficial to use smoothed quantities for the gradient and variance when estimating $\gamma_i$ and $s_i$. For this reason, we use bias-corrected exponential moving averages $\chi_i$ and $\xi_i$ in place of $g_i$ and $S_i$, respectively, when implementing equations \eqref{eq:gain_per_shot} and \eqref{eq:shots_iCANS}. These exponential moving averages introduce a new parameter, $\mu$, which controls the degree of smoothing and is bounded between $0$ and $1$. Since the update step is independent of this smoothing, we often find it beneficial to choose $\mu$ close to $1$ to achieve a steady progression of $s_i$'s. Finally, we also add a regularizing parameter $b$ to the denominators of lines 13, 16, and 20 of Algorthim~\ref{alg:iCANS} for numerical stability. By multiplying $b$ by $\mu^k$ and choosing $b$ to be small, the bias from this regularizing parameter begins small and exponentially decays as the algorithm progresses.

\begin{figure}[t]
    \centering
    \includegraphics[width=.8\columnwidth]{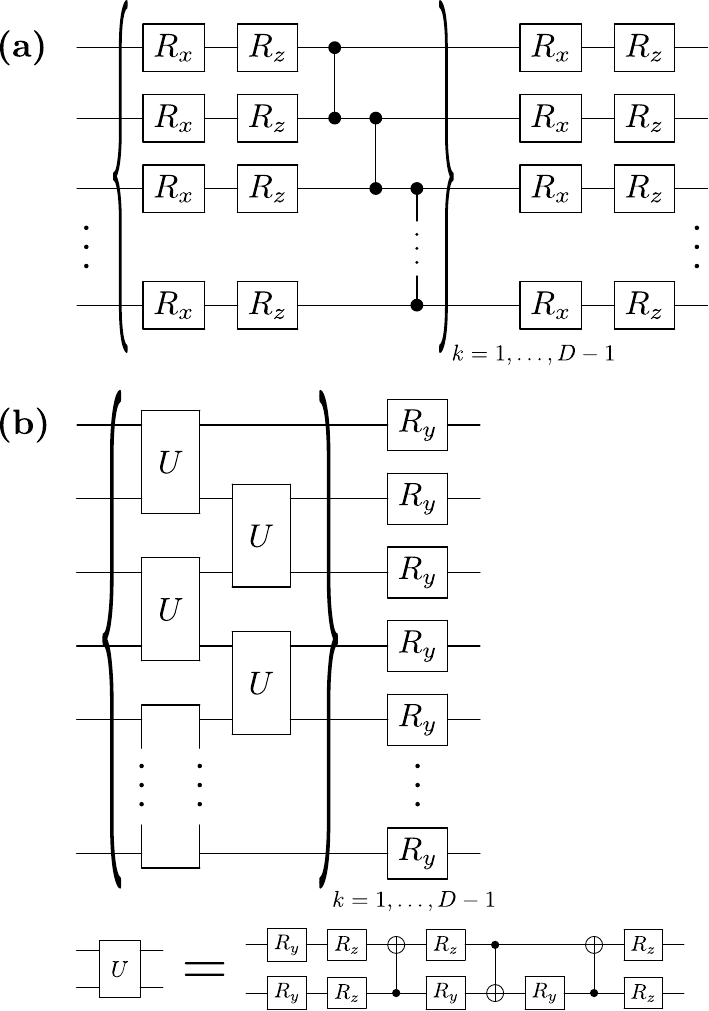}
    \caption{The quantum circuit diagram for the ansatzes we used to construct the unitary operators $U(\fattheta)$ in our implementations. The angles in each rotation gate (denoted as $R_j$, where $j$ denotes the axis being rotated about) are varied independently.  Panel \textbf{a} shows the ansatz used in the compiling and Heisenberg spin chain VQE task, and we note that this is the same ansatz used in Ref.~\cite{nakanishi2019}. Panel \textbf{b} shows the ansatz used when doing the size scaling comparison with the Ising spin chain VQE task.}
    \label{fig:ansatz}
\end{figure}
\begin{figure*}[t]
    \centering
    \includegraphics[width=1.8\columnwidth]{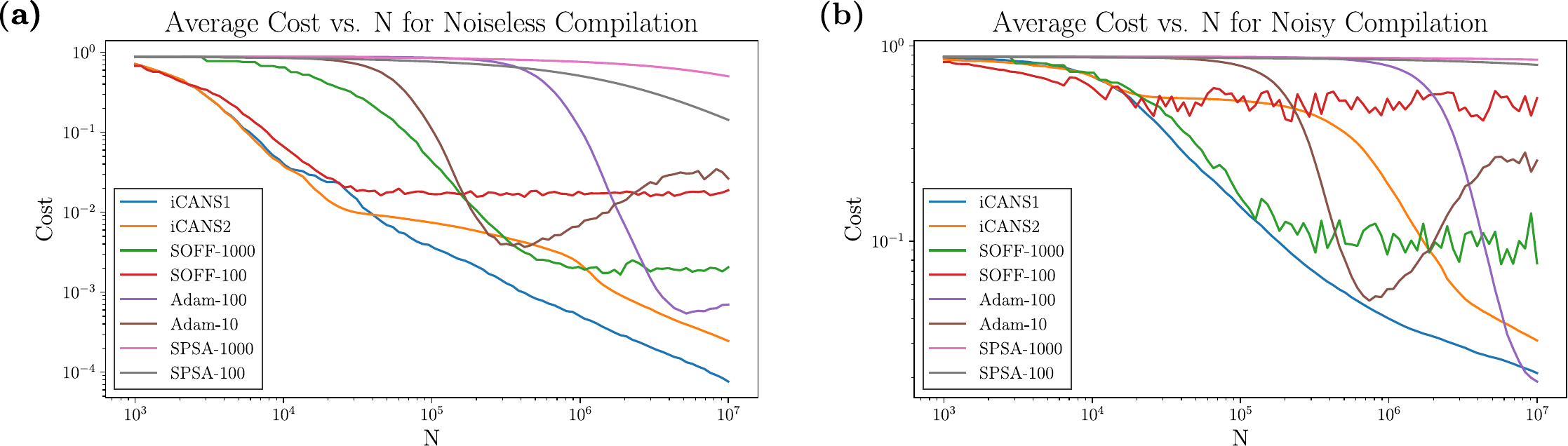}
    \caption{Comparison of performance for the compilation task across one hundred random target states and initial starts. As mentioned in the text, we denote algorithm A with $s$ shots per operator measurement as A-$s$.  Panels \textbf{a} and \textbf{b} show the average cost value  attained as a function of the total number of shots ($N$) expended for the noiseless and noisy cases, respectively.}
    \label{fig:compiling}
    
\end{figure*}

\begin{figure*}[!t]
    \centering
    \includegraphics[width=1.8\columnwidth]{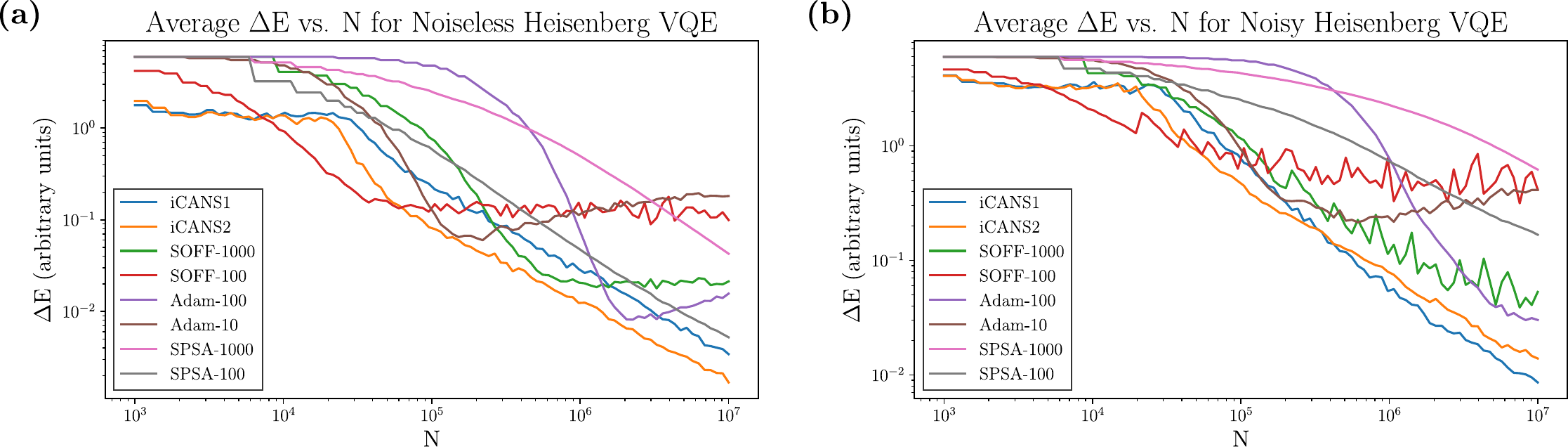}%
    \caption{Comparison of performance for the Heisenberg spin chain VQE task across one hundred random starts. Panels \textbf{a} and \textbf{b} show the average $\Delta E$ value (i.e. energy above the ground state energy) attained as a function of the total number of shots ($N$) expended for the noiseless and noisy cases, respectively.}
    \label{fig:vqe}
\end{figure*}

\section{Implementations} \label{sec:Implementations}

In order to compare the performance of iCANS1 and iCANS2 to established methods, we consider two optimization tasks: variational quantum compiling with a fixed input state~\cite{Khatri2019quantumassisted, jones2018quantum, heya2018variational, sharma2019noise} and a variational quantum eigensolver (VQE)~\cite{peruzzo2014VQE} for a Heisenberg spin chain.

For our experiments we set the iCANS hyperparameters as $\alpha = 0.1$, $\mu = 0.99$, and $b = 10^{-6}$, except for the case of the system size scaling comparison. For that case, since the Lipschitz constant $L$ grows linearly with the system size, leaving $\alpha = 0.1$ leads to $\alpha>2/L$ for larger systems, which is invalid for iCANS. We therefore chose $\alpha = 1/L$ for the different length Ising spin chains we consider below.

For the other algorithms we compare to, we will denote the number of shots per operator measurement as $s$. We will denote algorithm A with $s$ shots per operator measurement as A-$s$ (e.g., SOFF with $s=1000$ is denoted SOFF-1000). We also note that in the figures and tables below we show the analytical cost and energies that one could achieve with the parameters that the optimizers output, i.e., without hardware noise or shot noise. The optimizers did have to contend with finite statistics and, where indicated, hardware noise to find those parameters.

In addition to the fixed number of shots they use, the other algorithms we compare to also come with other hyperparameters, which were chosen empirically in an attempt to get the best performance from each. For Adam we used a learning rate of $\alpha=0.1$ along with the momentum parameter values of $\beta_1=0.9$ and $\beta_2=0.999$. For SPSA, we found that the default parameters were the best among those that we tried, and thus we set $A$ to be a tenth of the total number of allowed iterations, $\beta=0.602$, and $\gamma=0.101$.

\begin{table*}[!ht]
    \begin{minipage}{.5\linewidth}
      \centering
      
      \caption{Noiseless Compilation Average Cost Values}\label{tbl:Noiseless_comp}
      
\begin{adjustbox}{width=.95\columnwidth,center}
        \begin{tabular}{|l||r|r|r|r|r|}
        \hline
        \# Shots ($N$) & $10^3$       & $10^4$       & $10^5$       &  $10^6$ &  $10^7$   \\ \hhline{|=||=|=|=|=|=|}
 iCANS1    & 0.7140 & 0.0410 & \textbf{0.0037} & \textbf{0.0005} & \textbf{0.0001} \\\hline
 iCANS2    & 0.7149 & \textbf{0.0386} & 0.0074 & 0.0022 & 0.0002 \\\hline
 SOFF-1000 & X                  & 0.6483 & 0.0430 & 0.0020 & 0.0021 \\\hline
 SOFF-100  & \textbf{0.6761} & 0.0652 & 0.0185 & 0.0162 & 0.0164 \\\hline
 Adam-100  & 0.8814 & 0.8807 & 0.8578 & 0.1113 & 0.0007 \\\hline
 Adam-10   & 0.8807 & 0.8576 & 0.1108 & 0.0072 & 0.0289 \\\hline
 SPSA-1000 & X                  & 0.8693 & 0.8426 & 0.7587 & 0.5009 \\\hline
 SPSA-100  & 0.8719 & 0.8455 & 0.7625 & 0.5077 & 0.1428 \\\hline
        \end{tabular}
\end{adjustbox}
    \end{minipage}%
    \begin{minipage}{.5\linewidth}
      \centering
      \caption{Noisy Compilation Average Cost Values}\label{tbl:Noisy_comp}
\begin{adjustbox}{width=.95\columnwidth,center}
        \begin{tabular}{|l||r|r|r|r|r|}
        \hline
         \# Shots ($N$) & $10^3$       & $10^4$       & $10^5$       &  $10^6$ &  $10^7$   \\ \hhline{|=||=|=|=|=|=|}
 iCANS1    & 0.8711 & 0.6984 & \textbf{0.1498} & \textbf{0.0402} & \textbf{0.0211} \\\hline
 iCANS2    & 0.8527 & 0.6982 & 0.5236 & 0.1926 & 0.0310 \\\hline
 SOFF-1000 & X                  & 0.7302 & 0.1634 & 0.1157 & 0.0912 \\\hline
 SOFF-100  & \textbf{0.8272} & \textbf{0.6109} & 0.5506 & 0.4337 & 0.5740 \\\hline
 Adam-100  & 0.8814  & 0.8813 & 0.8791 & 0.7911 & 0.0191 \\\hline
 Adam-10   & 0.8813 & 0.8790 & 0.7918 & 0.0556 & 0.2583 \\\hline
 SPSA-1000 & X                  & 0.8775 & 0.8744 & 0.8679 & 0.8504 \\\hline
 SPSA-100  & 0.8761  & 0.8732 & 0.8669 & 0.8503 & 0.8006 \\\hline
        \end{tabular}
\end{adjustbox}
    \end{minipage} 
    \begin{minipage}{.5\linewidth}
      \centering
      \caption{Noiseless VQE Average $\Delta$ Energies}\label{tbl:Noiseless_vqe}
\begin{adjustbox}{width=.95\columnwidth,center}
        \begin{tabular}{|l||r|r|r|r|r|}
        \hline
         \# Shots ($N$)  & $10^3$       & $10^4$       & $10^5$       &  $10^6$ &  $10^7$   \\ \hhline{|=||=|=|=|=|=|}
iCANS1     & \textbf{1.7732}  &     1.3746 &      0.2478 &       0.0290 &        0.0034 \\\hline
 iCANS2    & 1.9755 &     1.3813 &      \textbf{0.0831} &       \textbf{0.0124} &        \textbf{0.0017} \\\hline
 SOFF-1000  & X                  &     4.0944 &      0.7970 &       0.0207 &        0.0213 \\\hline
 SOFF-100   & 4.2024  &     \textbf{0.9666} &      0.1221 &       0.1536 &        0.0993 \\\hline
 Adam-100  & 5.9849  &     5.9849 &      4.8078 &       0.0818 &        0.0157 \\\hline
 Adam-10    & 5.9849  &     4.8293 &      0.1431 &       0.1126 &        0.1816 \\\hline
 SPSA-1000 & X                  &     5.1937 &      2.5710 &       0.5067 &        0.0426 \\\hline
 SPSA-100  & 5.9849  &     3.2301 &      0.6240 &       0.0485 &        0.0052 \\\hline
        \end{tabular}
\end{adjustbox}
    \end{minipage}%
    \begin{minipage}{.5\linewidth}
      \centering
      \caption{Noisy VQE Average $\Delta$ Energies}\label{tbl:Noisy_vqe}
\begin{adjustbox}{width=.95\columnwidth,center}
        \begin{tabular}{|l||r|r|r|r|r|}
        \hline
         \# Shots ($N$) & $10^3$       & $10^4$       & $10^5$       &  $10^6$ &  $10^7$   \\ \hhline{|=||=|=|=|=|=|}
 iCANS1    & 4.1202  &     3.3035 &      0.8363 &       \textbf{0.0540} &        \textbf{0.0086} \\\hline
 iCANS2    & \textbf{4.0782} &     3.1897 &      \textbf{0.4918} &       0.0796 &        0.0139 \\\hline
 SOFF-1000 & X                 &     4.3083 &      1.1973 &       0.1137 &        0.0531 \\\hline
 SOFF-100  & 4.6518 &     \textbf{2.1337} &      0.6384 &       0.7574 &        0.4090 \\\hline
 Adam-100  & 5.9382 &     5.9382 &      5.5313 &       0.8062 &        0.0301 \\\hline
 Adam-10   & 5.9849 &     5.5759 &      1.0113 &       0.2428 &        0.4119    \\\hline
 SPSA-1000 & X                 &     5.6286 &      4.3573 &       2.2965 &        0.6206 \\\hline
 SPSA-100  & 5.9849 &     4.7220 &      2.5740 &       0.7463 &        0.1674 \\\hline
        \end{tabular}
\end{adjustbox}
    \end{minipage} 
\end{table*}

\subsection{Variational Compiling with a Fixed Input State}

For our first optimization task, we follow~\cite{nakanishi2019} and consider as a benchmark the optimization of the following cost function:
\begin{equation}
     C=1-\left| \left<\boldsymbol{0} \right | U(\fattheta^*)^{\dagger}U(\fattheta)\left | \boldsymbol{0} \right> \right|^2
\end{equation}
where $\fattheta^*$ is a vector of fixed, randomly selected angles and $\fattheta$ is the vector of angles to be optimized over. For this problem, we construct the parametrized unitary operator $U(\fattheta)$ with the ansatz described in Fig.~\ref{fig:ansatz}(a), setting $n=3$ qubits and $D=6$. (As adding depth and thus more parameters increases the difficulty of the optimization task and amplifies the effect of the noise model, $D=6$ was chosen to increase the difficulty of the task although shorter depth ansatzes would work here.) We then simulate the optimization procedure with one hundred different random seeds (each of which generates a unique random $\fattheta^*$ and initial point) and a collection of different optimizers. The results for both the case of a noiseless simulator and the case of a simulator using the noise profile of IBM's Melbourne processor~\cite{IBMQ14}  are shown in Fig.~\ref{fig:compiling}. For the latter, we emphasize that this noise profile reflects the properties of real, currently available quantum hardware. In addition, the average costs obtained for each optimizer are listed in Tables \ref{tbl:Noiseless_comp} and \ref{tbl:Noisy_comp} with the best value found for each total number of shots expended $N$ shown in bold. Furthermore, see Appendix~\ref{App:Cumulative} for the cumulative probability distributions over cost values, which provide more information than the average cost values.

\afterpage{\FloatBarrier}

\subsection{VQE}

For our second optimization task, we follow~\cite{kandala2017} in considering the Heisenberg spin chain with wrapped boundary conditions and the Hamiltonian:
\begin{equation}
    H=J\sum_{<ij>}\left(X_iX_j+Y_iY_j+Z_iZ_j\right) + B\sum_i Z_i,
\end{equation}
where the $<>$ bracket denotes nearest-neighbor pairs. For the purpose of our comparison, we fix $J=1$ and $B=3$ and again consider the ansatz described in Fig.~\ref{fig:ansatz}(a). Running the comparison with $n=3$ qubits in a triangle and $D=6$ for the ansatz, we simulate running VQE with one hundred different random seeds and initial points, along with the same set of optimizers as in the benchmark case above. As before, the results for the both a noiseless and a noisy simulator (also using the IBM Melbourne processor's noise profile~\cite{IBMQ14}) are shown in Fig.~\ref{fig:vqe}. Again, the average energies obtained for each optimizer are listed in Tables \ref{tbl:Noiseless_vqe} and \ref{tbl:Noisy_vqe} with the best value found for each total number of shots expended $N$ shown in bold. In addition, see Appendix~\ref{App:Cumulative} for the cumulative probability distributions over energy values, which provide more information than the average energy values.

\subsection{Comparison of Scaling}

In order to compare the performance of iCANS to that of other optimizers when one scales up the number of qubits, we now consider VQE applied to Ising spin chains of differing lengths with open boundary conditions and the Hamiltonian:
\begin{equation}
    H = - \sum_{<ij>}Z_iZ_j -g\sum_i X_i,
\end{equation} 
where the $<>$ bracket again denotes nearest-neighbor pairs. In order to generate enough entanglement in the ground state to require a modest depth, we choose $g=1.5$ so that we are near but not at the critical point $g=1$. For this problem, we used the ansatz shown in Fig.~\ref{fig:ansatz}(b) with $D=3$ (two repetitions of the block shown in braces), as its performance for this problem was significantly better than the simple ansatz in Fig.~\ref{fig:ansatz}(a). 

\begin{figure*}[!t]
    \centering
    \includegraphics[width=2\columnwidth]{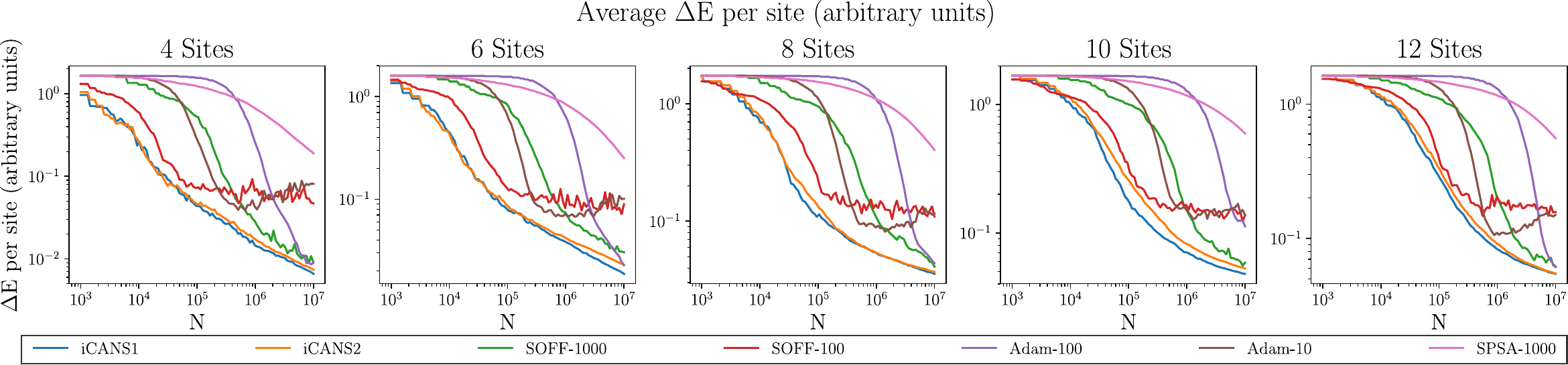}
    \caption{Comparison of performance for the Ising VQE task with different numbers of sites (i.e., qubits) without hardware noise. Each panel shows the average $\Delta E$ per site value attained as a function of the total number of shots expended ($N$) for each number of qubits. Each curve represents the average over ten random starts.}
    \label{fig:vqe_scaling}
\end{figure*}

The results for a noiseless simulator for 4, 6, 8, 10, and 12 qubit Ising spin chains are shown in Fig.~\ref{fig:vqe_scaling}.

\section{Discussion}
\label{sec:Discussion}
Here we report on the behavior of the various optimizers we studied. First we consider SOFF, which is the only optimizer studied here other than iCANS that was formulated specifically for VHQCAs. By leveraging analytical knowledge about the optimization landscape, SOFF's gradient-free method of making single parameter updates allows it to quickly train in low noise environments. However, the limit of the precision when fitting the analytical function with a finite number of shots means that SOFF hits a precision floor and cannot improve past that point. Additionally, hardware noise tends to distort the landscape in such a way that the analytical form no longer provides as good of a fit, making SOFF struggle more relative to the other optimizers considered. In the optimization tasks we looked at here, we found that SOFF was often competitive with iCANS shortly before hitting its precision floor, with SOFF-100 sometimes doing better for a brief interval early on. For example, SOFF-100 was the best performing optimizer for the compilation task with $N=10^3$ (noiseless and noisy) and $N=10^4$ (noisy only), as well as for the Heisenberg VQE with $N = 10^4$  (noiseless and noisy).

Adam was originally conceived in the context of machine learning and excels at optimizing in noisy environments. However, in our numerical studies we found that Adam suffered from an instability given the hyperparameters we chose and the number of shots we allowed at each partial derivative evaluation. This instability appears to enter later in the optimization when we are working with more shots, and it can be seen in the upturn of the curves in Figs.~\ref{fig:compiling}--\ref{fig:vqe_scaling}. For the case of the noisy compilation task, Adam-100 looks like it might just be reaching that instability at the end of the allowed shot budget, and slightly outperformed iCANS1 to be the best on average. We note that, similar to what was seen with SOFF, Adam was usually competitive with the iCANS methods before it reached the point where it stopped improving.
\afterpage{\FloatBarrier}

Unlike SOFF and Adam, SPSA did not seem to hit a point at which it stopped improving with shot budget, for the chosen hyperparameters. We note though that SPSA is the most sensitive to perturbations of the hyperparameters among the methods studied here and can become very unstable if they are incorrectly chosen. However, if one hits upon the correct hyperparameters, SPSA can be very effective. While for our cases we did not find SPSA outperforming iCANS, we note that for the noiseless Heisenberg VQE task, SPSA-100 was the most competitive with iCANS. 

Overall, we find that iCANS performed well on all optimization tasks considered, with either iCANS1 or iCANS2 usually providing the best result for a given total shot budget $N$. Even when scaling up the system size in the Ising model VQE task (see Fig.~\ref{fig:vqe_scaling}), we found that iCANS continued to outperform the other optimizers studied. We also note that empirically iCANS1 usually outperformed iCANS2. While iCANS2 provides a benefit by reducing the sensitivity to the input learning rate, so long as the learning rate is chosen well we expect that iCANS1 may tend to perform better.

We remark that while we do not report full results for RAdam~\cite{liu2019variance}, we found with preliminary results that it did not seem to provide a substantial improvement over the simpler Adam algorithm for our use cases. 
Similarly, we found that SOFF with the Anderson acceleration step proposed in~\cite{parrish2019} did not noticeably improve upon the performance of basic SOFF, and therefore the curves for this method are not shown.


We finally remark about the different performance for the various fixed-shot optimizers with different numbers of shots (e.g., Adam-10 versus Adam-100). This performance difference can be understood as a trade-off between reducing the statistical uncertainty and achieving more iterations before hitting the limit on the total number of shots. When few shots are used, many more iterations might be allowed but the update steps are much noisier, usually meaning that the optimizer can perform more quickly early on but then potentially hits an effective floor due to the precision. Increasing the number of shots will allow more precise updates and thus lowers the precision floor (if present) but means that far fewer iterations can be performed. This is the idea at the heart of iCANS. iCANS uses few shots early on and so achieves a period of noisy but fast descent, but then slows down and computes with greater and greater precision to continue making progress. This strategy allows for shot frugality as well as in principle removing such a precision floor for iCANS.

\section{Conclusions}

In order to bring about the promise of VHQCAs solving usefully large and complex problems on NISQ devices, one needs a way to perform the requisite optimizations efficiently. As the rate-limiting step of these optimizations will likely be the number of times one must prepare and measure quantum states, it will be important to have optimizers that are frugal in the number of times physical measurements must be performed on a quantum computer. 

In this work we introduced two versions of a measurement-frugal, noise-resilient optimizer tailored for VHQCAs. Both of the strategies we propose, iCANS1 and iCANS2, address measurement frugality by dynamically determining the number of measurements needed for each partial derivative of each step in a gradient descent. iCANS1 is the more aggressive version, always taking the same learning rate, while iCANS2 is more cautious and limits the learning rate for steps so that the expected gain is always guaranteed to be positive. Our numerical results indicate that these optimizers may perform comparably or better than other state-of-the-art optimizers. The performance compares especially well in the presence of realistic hardware noise.

iCANS has already found use in the very recent VHQCA literature~\cite{sharma2019noise}. Furthermore, after our article was originally posted, a related study of stochastic gradient descent for VHQCAs found that small shot counts can provide rapid improvement in early stages of training~\cite{sweke2019stochastic}, which provides further motivation for iCANS.

One potential direction for future work is exploring the possibility of extending our frugal adaptive approach to non-gradient methods, such as SPSA.


\begin{acknowledgements}

JMK acknowledges support from the U.S. Department of Energy (DOE) through a quantum computing program sponsored by the Los Alamos National Laboratory (LANL) Information Science \& Technology Institute.  AA, LC, and PJC acknowledge support from the LDRD program at LANL. PJC also acknowledges support from the LANL ASC Beyond Moore's Law project. This work was also supported by the U.S. DOE, Office of Science, Office of Advanced Scientific Computing Research, under the Quantum Computing Application Teams program.

\end{acknowledgements}
\bibliographystyle{apsrev4-1mod}

\begin{thebibliography}{59}%
\makeatletter
\providecommand \@ifxundefined [1]{%
 \@ifx{#1\undefined}
}%
\providecommand \@ifnum [1]{%
 \ifnum #1\expandafter \@firstoftwo
 \else \expandafter \@secondoftwo
 \fi
}%
\providecommand \@ifx [1]{%
 \ifx #1\expandafter \@firstoftwo
 \else \expandafter \@secondoftwo
 \fi
}%
\providecommand \natexlab [1]{#1}%
\providecommand \enquote  [1]{``#1''}%
\providecommand \bibnamefont  [1]{#1}%
\providecommand \bibfnamefont [1]{#1}%
\providecommand \citenamefont [1]{#1}%
\providecommand \href@noop [0]{\@secondoftwo}%
\providecommand \href [0]{\begingroup \@sanitize@url \@href}%
\providecommand \@href[1]{\@@startlink{#1}\@@href}%
\providecommand \@@href[1]{\endgroup#1\@@endlink}%
\providecommand \@sanitize@url [0]{\catcode `\\12\catcode `\$12\catcode
  `\&12\catcode `\#12\catcode `\^12\catcode `\_12\catcode `\%12\relax}%
\providecommand \@@startlink[1]{}%
\providecommand \@@endlink[0]{}%
\providecommand \url  [0]{\begingroup\@sanitize@url \@url }%
\providecommand \@url [1]{\endgroup\@href {#1}{\urlprefix }}%
\providecommand \urlprefix  [0]{URL }%
\providecommand \Eprint [0]{\href }%
\providecommand \doibasemod [0]{http://dx.doi.org/}%
\providecommand \selectlanguage [0]{\@gobble}%
\providecommand \bibinfo  [0]{\@secondoftwo}%
\providecommand \bibfield  [0]{\@secondoftwo}%
\providecommand \translation [1]{[#1]}%
\providecommand \BibitemOpen [0]{}%
\providecommand \bibitemStop [0]{}%
\providecommand \bibitemNoStop [0]{.\EOS\space}%
\providecommand \EOS [0]{\spacefactor3000\relax}%
\providecommand \BibitemShut  [1]{\csname bibitem#1\endcsname}%
\let\auto@bib@innerbib\@empty
\bibitem [{\citenamefont {Preskill}(2018)}]{preskill2018quantum}%
  \BibitemOpen
  \bibfield  {author} {\bibinfo {author} {\bibfnamefont {J.}~\bibnamefont
  {Preskill}},\ }\bibfield  {title} {\emph {\bibinfo {title} {Quantum computing
  in the {NISQ} era and beyond},\ }}\href {\doibasemod
  10.22331/q-2018-08-06-79} {\bibfield  {journal} {\bibinfo  {journal}
  {Quantum}\ }\textbf {\bibinfo {volume} {2}},\ \bibinfo {pages} {79} (\bibinfo
  {year} {2018})}\BibitemShut {NoStop}%
\bibitem [{\citenamefont {McClean}\ \emph {et~al.}(2016)\citenamefont
  {McClean}, \citenamefont {Romero}, \citenamefont {Babbush},\ and\
  \citenamefont {Aspuru-Guzik}}]{mcclean2016theory}%
  \BibitemOpen
  \bibfield  {author} {\bibinfo {author} {\bibfnamefont {J.~R.}\ \bibnamefont
  {McClean}}, \bibinfo {author} {\bibfnamefont {J.}~\bibnamefont {Romero}},
  \bibinfo {author} {\bibfnamefont {R.}~\bibnamefont {Babbush}}, \ and\
  \bibinfo {author} {\bibfnamefont {A.}~\bibnamefont {Aspuru-Guzik}},\
  }\bibfield  {title} {\emph {\bibinfo {title} {The theory of variational
  hybrid quantum-classical algorithms},\ }}\href {\doibasemod
  10.1088/1367-2630/18/2/023023} {\bibfield  {journal} {\bibinfo  {journal}
  {New Journal of Physics}\ }\textbf {\bibinfo {volume} {18}},\ \bibinfo
  {pages} {023023} (\bibinfo {year} {2016})}\BibitemShut {NoStop}%
\bibitem [{\citenamefont {Peruzzo}\ \emph {et~al.}(2014)\citenamefont
  {Peruzzo}, \citenamefont {McClean}, \citenamefont {Shadbolt}, \citenamefont
  {Yung}, \citenamefont {Zhou}, \citenamefont {Love}, \citenamefont
  {Aspuru-Guzik},\ and\ \citenamefont {O'brien}}]{peruzzo2014VQE}%
  \BibitemOpen
  \bibfield  {author} {\bibinfo {author} {\bibfnamefont {A.}~\bibnamefont
  {Peruzzo}}, \bibinfo {author} {\bibfnamefont {J.}~\bibnamefont {McClean}},
  \bibinfo {author} {\bibfnamefont {P.}~\bibnamefont {Shadbolt}}, \bibinfo
  {author} {\bibfnamefont {M.-H.}\ \bibnamefont {Yung}}, \bibinfo {author}
  {\bibfnamefont {X.-Q.}\ \bibnamefont {Zhou}}, \bibinfo {author}
  {\bibfnamefont {P.~J.}\ \bibnamefont {Love}}, \bibinfo {author}
  {\bibfnamefont {A.}~\bibnamefont {Aspuru-Guzik}}, \ and\ \bibinfo {author}
  {\bibfnamefont {J.~L.}\ \bibnamefont {O'brien}},\ }\bibfield  {title} {\emph
  {\bibinfo {title} {A variational eigenvalue solver on a photonic quantum
  processor},\ }}\href {\doibasemod 10.1038/ncomms5213} {\bibfield  {journal}
  {\bibinfo  {journal} {Nature Communications}\ }\textbf {\bibinfo {volume}
  {5}},\ \bibinfo {pages} {4213} (\bibinfo {year} {2014})}\BibitemShut
  {NoStop}%
\bibitem [{\citenamefont {Farhi}\ \emph {et~al.}(2014)\citenamefont {Farhi},
  \citenamefont {Goldstone},\ and\ \citenamefont {Gutmann}}]{farhi2014QAOA}%
  \BibitemOpen
  \bibfield  {author} {\bibinfo {author} {\bibfnamefont {E.}~\bibnamefont
  {Farhi}}, \bibinfo {author} {\bibfnamefont {J.}~\bibnamefont {Goldstone}}, \
  and\ \bibinfo {author} {\bibfnamefont {S.}~\bibnamefont {Gutmann}},\
  }\bibfield  {title} {\emph {\bibinfo {title} {A quantum approximate
  optimization algorithm},\ }}\href {https://arxiv.org/abs/1411.4028}
  {\bibfield  {journal} {\bibinfo  {journal} {arXiv:1411.4028}\ } (\bibinfo
  {year} {2014})}\BibitemShut {NoStop}%
\bibitem [{\citenamefont {Johnson}\ \emph {et~al.}(2017)\citenamefont
  {Johnson}, \citenamefont {Romero}, \citenamefont {Olson}, \citenamefont
  {Cao},\ and\ \citenamefont {Aspuru-Guzik}}]{johnson2017qvector}%
  \BibitemOpen
  \bibfield  {author} {\bibinfo {author} {\bibfnamefont {P.~D.}\ \bibnamefont
  {Johnson}}, \bibinfo {author} {\bibfnamefont {J.}~\bibnamefont {Romero}},
  \bibinfo {author} {\bibfnamefont {J.}~\bibnamefont {Olson}}, \bibinfo
  {author} {\bibfnamefont {Y.}~\bibnamefont {Cao}}, \ and\ \bibinfo {author}
  {\bibfnamefont {A.}~\bibnamefont {Aspuru-Guzik}},\ }\bibfield  {title} {\emph
  {\bibinfo {title} {{QVECTOR}: an algorithm for device-tailored quantum error
  correction},\ }}\href {https://arxiv.org/abs/1711.02249} {\bibfield
  {journal} {\bibinfo  {journal} {arXiv:1711.02249}\ } (\bibinfo {year}
  {2017})}\BibitemShut {NoStop}%
\bibitem [{\citenamefont {Romero}\ \emph {et~al.}(2017)\citenamefont {Romero},
  \citenamefont {Olson},\ and\ \citenamefont
  {Aspuru-Guzik}}]{romero2017quantum}%
  \BibitemOpen
  \bibfield  {author} {\bibinfo {author} {\bibfnamefont {J.}~\bibnamefont
  {Romero}}, \bibinfo {author} {\bibfnamefont {J.~P.}\ \bibnamefont {Olson}}, \
  and\ \bibinfo {author} {\bibfnamefont {A.}~\bibnamefont {Aspuru-Guzik}},\
  }\bibfield  {title} {\emph {\bibinfo {title} {Quantum autoencoders for
  efficient compression of quantum data},\ }}\href {\doibasemod
  10.1088/2058-9565/aa8072} {\bibfield  {journal} {\bibinfo  {journal} {Quantum
  Science and Technology}\ }\textbf {\bibinfo {volume} {2}},\ \bibinfo {pages}
  {045001} (\bibinfo {year} {2017})}\BibitemShut {NoStop}%
\bibitem [{\citenamefont {LaRose}\ \emph {et~al.}(2019)\citenamefont {LaRose},
  \citenamefont {Tikku}, \citenamefont {O'Neel-Judy}, \citenamefont {Cincio},\
  and\ \citenamefont {Coles}}]{larose2018}%
  \BibitemOpen
  \bibfield  {author} {\bibinfo {author} {\bibfnamefont {R.}~\bibnamefont
  {LaRose}}, \bibinfo {author} {\bibfnamefont {A.}~\bibnamefont {Tikku}},
  \bibinfo {author} {\bibfnamefont {{\'E}.}~\bibnamefont {O'Neel-Judy}},
  \bibinfo {author} {\bibfnamefont {L.}~\bibnamefont {Cincio}}, \ and\ \bibinfo
  {author} {\bibfnamefont {P.~J.}\ \bibnamefont {Coles}},\ }\bibfield  {title}
  {\emph {\bibinfo {title} {Variational quantum state diagonalization},\
  }}\href {\doibasemod 10.1038/s41534-019-0167-6} {\bibfield  {journal}
  {\bibinfo  {journal} {npj Quantum Information}\ }\textbf {\bibinfo {volume}
  {5}},\ \bibinfo {pages} {57} (\bibinfo {year} {2019})}\BibitemShut {NoStop}%
\bibitem [{\citenamefont {Arrasmith}\ \emph {et~al.}(2019)\citenamefont
  {Arrasmith}, \citenamefont {Cincio}, \citenamefont {Sornborger},
  \citenamefont {Zurek},\ and\ \citenamefont
  {Coles}}]{arrasmith2019variational}%
  \BibitemOpen
  \bibfield  {author} {\bibinfo {author} {\bibfnamefont {A.}~\bibnamefont
  {Arrasmith}}, \bibinfo {author} {\bibfnamefont {L.}~\bibnamefont {Cincio}},
  \bibinfo {author} {\bibfnamefont {A.~T.}\ \bibnamefont {Sornborger}},
  \bibinfo {author} {\bibfnamefont {W.~H.}\ \bibnamefont {Zurek}}, \ and\
  \bibinfo {author} {\bibfnamefont {P.~J.}\ \bibnamefont {Coles}},\ }\bibfield
  {title} {\emph {\bibinfo {title} {Variational consistent histories as a
  hybrid algorithm for quantum foundations},\ }}\href {\doibasemod
  10.1038/s41467-019-11417-0} {\bibfield  {journal} {\bibinfo  {journal}
  {Nature Communications}\ }\textbf {\bibinfo {volume} {10}},\ \bibinfo {pages}
  {3438} (\bibinfo {year} {2019})}\BibitemShut {NoStop}%
\bibitem [{\citenamefont {Cerezo}\ \emph {et~al.}(2020)\citenamefont {Cerezo},
  \citenamefont {Poremba}, \citenamefont {Cincio},\ and\ \citenamefont
  {Coles}}]{cerezo2019variational}%
  \BibitemOpen
  \bibfield  {author} {\bibinfo {author} {\bibfnamefont {M.}~\bibnamefont
  {Cerezo}}, \bibinfo {author} {\bibfnamefont {A.}~\bibnamefont {Poremba}},
  \bibinfo {author} {\bibfnamefont {L.}~\bibnamefont {Cincio}}, \ and\ \bibinfo
  {author} {\bibfnamefont {P.~J.}\ \bibnamefont {Coles}},\ }\bibfield  {title}
  {\emph {\bibinfo {title} {Variational quantum fidelity estimation},\ }}\href
  {\doibasemod 10.22331/q-2020-03-26-248} {\bibfield  {journal} {\bibinfo
  {journal} {Quantum}\ }\textbf {\bibinfo {volume} {4}},\ \bibinfo {pages}
  {248} (\bibinfo {year} {2020})}\BibitemShut {NoStop}%
\bibitem [{\citenamefont {Jones}\ \emph {et~al.}(2019)\citenamefont {Jones},
  \citenamefont {Endo}, \citenamefont {McArdle}, \citenamefont {Yuan},\ and\
  \citenamefont {Benjamin}}]{jones2019variational}%
  \BibitemOpen
  \bibfield  {author} {\bibinfo {author} {\bibfnamefont {T.}~\bibnamefont
  {Jones}}, \bibinfo {author} {\bibfnamefont {S.}~\bibnamefont {Endo}},
  \bibinfo {author} {\bibfnamefont {S.}~\bibnamefont {McArdle}}, \bibinfo
  {author} {\bibfnamefont {X.}~\bibnamefont {Yuan}}, \ and\ \bibinfo {author}
  {\bibfnamefont {S.~C.}\ \bibnamefont {Benjamin}},\ }\bibfield  {title} {\emph
  {\bibinfo {title} {Variational quantum algorithms for discovering hamiltonian
  spectra},\ }}\href {\doibasemod 10.1103/PhysRevA.99.062304} {\bibfield
  {journal} {\bibinfo  {journal} {Physical Review A}\ }\textbf {\bibinfo
  {volume} {99}},\ \bibinfo {pages} {062304} (\bibinfo {year}
  {2019})}\BibitemShut {NoStop}%
\bibitem [{\citenamefont {Yuan}\ \emph {et~al.}(2019)\citenamefont {Yuan},
  \citenamefont {Endo}, \citenamefont {Zhao}, \citenamefont {Li},\ and\
  \citenamefont {Benjamin}}]{yuan2018theory}%
  \BibitemOpen
  \bibfield  {author} {\bibinfo {author} {\bibfnamefont {X.}~\bibnamefont
  {Yuan}}, \bibinfo {author} {\bibfnamefont {S.}~\bibnamefont {Endo}}, \bibinfo
  {author} {\bibfnamefont {Q.}~\bibnamefont {Zhao}}, \bibinfo {author}
  {\bibfnamefont {Y.}~\bibnamefont {Li}}, \ and\ \bibinfo {author}
  {\bibfnamefont {S.~C.}\ \bibnamefont {Benjamin}},\ }\bibfield  {title} {\emph
  {\bibinfo {title} {Theory of variational quantum simulation},\ }}\href
  {\doibasemod 10.22331/q-2019-10-07-191} {\bibfield  {journal} {\bibinfo
  {journal} {Quantum}\ }\textbf {\bibinfo {volume} {3}},\ \bibinfo {pages}
  {191} (\bibinfo {year} {2019})}\BibitemShut {NoStop}%
\bibitem [{\citenamefont {Li}\ and\ \citenamefont
  {Benjamin}(2017)}]{li2017efficient}%
  \BibitemOpen
  \bibfield  {author} {\bibinfo {author} {\bibfnamefont {Y.}~\bibnamefont
  {Li}}\ and\ \bibinfo {author} {\bibfnamefont {S.~C.}\ \bibnamefont
  {Benjamin}},\ }\bibfield  {title} {\emph {\bibinfo {title} {Efficient
  variational quantum simulator incorporating active error minimization},\
  }}\href {\doibasemod 10.1103/PhysRevX.7.021050} {\bibfield  {journal}
  {\bibinfo  {journal} {Physical Review X}\ }\textbf {\bibinfo {volume} {7}},\
  \bibinfo {pages} {021050} (\bibinfo {year} {2017})}\BibitemShut {NoStop}%
\bibitem [{\citenamefont {Kokail}\ \emph {et~al.}(2019)\citenamefont {Kokail},
  \citenamefont {Maier}, \citenamefont {van Bijnen}, \citenamefont {Brydges},
  \citenamefont {Joshi}, \citenamefont {Jurcevic}, \citenamefont {Muschik},
  \citenamefont {Silvi}, \citenamefont {Blatt}, \citenamefont {Roos} \emph
  {et~al.}}]{kokail2019self}%
  \BibitemOpen
  \bibfield  {author} {\bibinfo {author} {\bibfnamefont {C.}~\bibnamefont
  {Kokail}}, \bibinfo {author} {\bibfnamefont {C.}~\bibnamefont {Maier}},
  \bibinfo {author} {\bibfnamefont {R.}~\bibnamefont {van Bijnen}}, \bibinfo
  {author} {\bibfnamefont {T.}~\bibnamefont {Brydges}}, \bibinfo {author}
  {\bibfnamefont {M.}~\bibnamefont {Joshi}}, \bibinfo {author} {\bibfnamefont
  {P.}~\bibnamefont {Jurcevic}}, \bibinfo {author} {\bibfnamefont
  {C.}~\bibnamefont {Muschik}}, \bibinfo {author} {\bibfnamefont
  {P.}~\bibnamefont {Silvi}}, \bibinfo {author} {\bibfnamefont
  {R.}~\bibnamefont {Blatt}}, \bibinfo {author} {\bibfnamefont
  {C.}~\bibnamefont {Roos}},  \emph {et~al.},\ }\bibfield  {title} {\emph
  {\bibinfo {title} {Self-verifying variational quantum simulation of lattice
  models},\ }}\href {\doibasemod 10.1038/s41586-019-1177-4} {\bibfield
  {journal} {\bibinfo  {journal} {Nature}\ }\textbf {\bibinfo {volume} {569}},\
  \bibinfo {pages} {355} (\bibinfo {year} {2019})}\BibitemShut {NoStop}%
\bibitem [{\citenamefont {Khatri}\ \emph {et~al.}(2019)\citenamefont {Khatri},
  \citenamefont {LaRose}, \citenamefont {Poremba}, \citenamefont {Cincio},
  \citenamefont {Sornborger},\ and\ \citenamefont
  {Coles}}]{Khatri2019quantumassisted}%
  \BibitemOpen
  \bibfield  {author} {\bibinfo {author} {\bibfnamefont {S.}~\bibnamefont
  {Khatri}}, \bibinfo {author} {\bibfnamefont {R.}~\bibnamefont {LaRose}},
  \bibinfo {author} {\bibfnamefont {A.}~\bibnamefont {Poremba}}, \bibinfo
  {author} {\bibfnamefont {L.}~\bibnamefont {Cincio}}, \bibinfo {author}
  {\bibfnamefont {A.~T.}\ \bibnamefont {Sornborger}}, \ and\ \bibinfo {author}
  {\bibfnamefont {P.~J.}\ \bibnamefont {Coles}},\ }\bibfield  {title} {\emph
  {\bibinfo {title} {Quantum-assisted quantum compiling},\ }}\href {\doibasemod
  10.22331/q-2019-05-13-140} {\bibfield  {journal} {\bibinfo  {journal}
  {{Quantum}}\ }\textbf {\bibinfo {volume} {3}},\ \bibinfo {pages} {140}
  (\bibinfo {year} {2019})}\BibitemShut {NoStop}%
\bibitem [{\citenamefont {Jones}\ and\ \citenamefont
  {Benjamin}(2018)}]{jones2018quantum}%
  \BibitemOpen
  \bibfield  {author} {\bibinfo {author} {\bibfnamefont {T.}~\bibnamefont
  {Jones}}\ and\ \bibinfo {author} {\bibfnamefont {S.~C.}\ \bibnamefont
  {Benjamin}},\ }\bibfield  {title} {\emph {\bibinfo {title} {Quantum
  compilation and circuit optimisation via energy dissipation},\ }}\href
  {https://arxiv.org/abs/1811.03147} {\bibfield  {journal} {\bibinfo  {journal}
  {arXiv:1811.03147}\ } (\bibinfo {year} {2018})}\BibitemShut {NoStop}%
\bibitem [{\citenamefont {Heya}\ \emph {et~al.}(2018)\citenamefont {Heya},
  \citenamefont {Suzuki}, \citenamefont {Nakamura},\ and\ \citenamefont
  {Fujii}}]{heya2018variational}%
  \BibitemOpen
  \bibfield  {author} {\bibinfo {author} {\bibfnamefont {K.}~\bibnamefont
  {Heya}}, \bibinfo {author} {\bibfnamefont {Y.}~\bibnamefont {Suzuki}},
  \bibinfo {author} {\bibfnamefont {Y.}~\bibnamefont {Nakamura}}, \ and\
  \bibinfo {author} {\bibfnamefont {K.}~\bibnamefont {Fujii}},\ }\bibfield
  {title} {\emph {\bibinfo {title} {Variational quantum gate optimization},\
  }}\href {https://arxiv.org/abs/1810.12745} {\bibfield  {journal} {\bibinfo
  {journal} {arXiv:1810.12745}\ } (\bibinfo {year} {2018})}\BibitemShut
  {NoStop}%
\bibitem [{\citenamefont {Endo}\ \emph {et~al.}(2018)\citenamefont {Endo},
  \citenamefont {Li}, \citenamefont {Benjamin},\ and\ \citenamefont
  {Yuan}}]{endo2018variational}%
  \BibitemOpen
  \bibfield  {author} {\bibinfo {author} {\bibfnamefont {S.}~\bibnamefont
  {Endo}}, \bibinfo {author} {\bibfnamefont {Y.}~\bibnamefont {Li}}, \bibinfo
  {author} {\bibfnamefont {S.}~\bibnamefont {Benjamin}}, \ and\ \bibinfo
  {author} {\bibfnamefont {X.}~\bibnamefont {Yuan}},\ }\bibfield  {title}
  {\emph {\bibinfo {title} {Variational quantum simulation of general
  processes},\ }}\href {https://arxiv.org/abs/1812.08778} {\bibfield  {journal}
  {\bibinfo  {journal} {arXiv:1812.08778}\ } (\bibinfo {year}
  {2018})}\BibitemShut {NoStop}%
\bibitem [{\citenamefont {Sharma}\ \emph {et~al.}(2020)\citenamefont {Sharma},
  \citenamefont {Khatri}, \citenamefont {Cerezo},\ and\ \citenamefont
  {Coles}}]{sharma2019noise}%
  \BibitemOpen
  \bibfield  {author} {\bibinfo {author} {\bibfnamefont {K.}~\bibnamefont
  {Sharma}}, \bibinfo {author} {\bibfnamefont {S.}~\bibnamefont {Khatri}},
  \bibinfo {author} {\bibfnamefont {M.}~\bibnamefont {Cerezo}}, \ and\ \bibinfo
  {author} {\bibfnamefont {P.}~\bibnamefont {Coles}},\ }\bibfield  {title}
  {\emph {\bibinfo {title} {Noise resilience of variational quantum
  compiling},\ }}\href {\doibasemod 10.1088/1367-2630/ab784c} {\bibfield
  {journal} {\bibinfo  {journal} {New Journal of Physics}\ } (\bibinfo {year}
  {2020}),\ 10.1088/1367-2630/ab784c}\BibitemShut {NoStop}%
\bibitem [{\citenamefont {Carolan}\ \emph {et~al.}(2019)\citenamefont
  {Carolan}, \citenamefont {Mosheni}, \citenamefont {Olson}, \citenamefont
  {Prabhu}, \citenamefont {Chen}, \citenamefont {Bunandar}, \citenamefont
  {Harris}, \citenamefont {Wong}, \citenamefont {Hochberg}, \citenamefont
  {Lloyd} \emph {et~al.}}]{carolan2019variational}%
  \BibitemOpen
  \bibfield  {author} {\bibinfo {author} {\bibfnamefont {J.}~\bibnamefont
  {Carolan}}, \bibinfo {author} {\bibfnamefont {M.}~\bibnamefont {Mosheni}},
  \bibinfo {author} {\bibfnamefont {J.~P.}\ \bibnamefont {Olson}}, \bibinfo
  {author} {\bibfnamefont {M.}~\bibnamefont {Prabhu}}, \bibinfo {author}
  {\bibfnamefont {C.}~\bibnamefont {Chen}}, \bibinfo {author} {\bibfnamefont
  {D.}~\bibnamefont {Bunandar}}, \bibinfo {author} {\bibfnamefont {N.~C.}\
  \bibnamefont {Harris}}, \bibinfo {author} {\bibfnamefont {F.~N.}\
  \bibnamefont {Wong}}, \bibinfo {author} {\bibfnamefont {M.}~\bibnamefont
  {Hochberg}}, \bibinfo {author} {\bibfnamefont {S.}~\bibnamefont {Lloyd}},
  \emph {et~al.},\ }\bibfield  {title} {\emph {\bibinfo {title} {Variational
  quantum unsampling on a quantum photonic processor},\ }}\href
  {https://arxiv.org/abs/1904.10463} {\bibfield  {journal} {\bibinfo  {journal}
  {arXiv:1904.10463}\ } (\bibinfo {year} {2019})}\BibitemShut {NoStop}%
\bibitem [{\citenamefont {Yoshioka}\ \emph {et~al.}(2019)\citenamefont
  {Yoshioka}, \citenamefont {Nakagawa}, \citenamefont {Mitarai},\ and\
  \citenamefont {Fujii}}]{yoshioka2019variational}%
  \BibitemOpen
  \bibfield  {author} {\bibinfo {author} {\bibfnamefont {N.}~\bibnamefont
  {Yoshioka}}, \bibinfo {author} {\bibfnamefont {Y.~O.}\ \bibnamefont
  {Nakagawa}}, \bibinfo {author} {\bibfnamefont {K.}~\bibnamefont {Mitarai}}, \
  and\ \bibinfo {author} {\bibfnamefont {K.}~\bibnamefont {Fujii}},\ }\bibfield
   {title} {\emph {\bibinfo {title} {Variational quantum algorithm for
  non-equilirium steady states},\ }}\href {https://arxiv.org/abs/1908.09836}
  {\bibfield  {journal} {\bibinfo  {journal} {arXiv:1908.09836}\ } (\bibinfo
  {year} {2019})}\BibitemShut {NoStop}%
\bibitem [{\citenamefont {Bravo-Prieto}\ \emph {et~al.}(2019)\citenamefont
  {Bravo-Prieto}, \citenamefont {LaRose}, \citenamefont {Cerezo}, \citenamefont
  {Subasi}, \citenamefont {Cincio},\ and\ \citenamefont
  {Coles}}]{bravo-prieto2019}%
  \BibitemOpen
  \bibfield  {author} {\bibinfo {author} {\bibfnamefont {C.}~\bibnamefont
  {Bravo-Prieto}}, \bibinfo {author} {\bibnamefont {LaRose}}, \bibinfo {author}
  {\bibfnamefont {M.}~\bibnamefont {Cerezo}}, \bibinfo {author} {\bibfnamefont
  {Y.}~\bibnamefont {Subasi}}, \bibinfo {author} {\bibfnamefont
  {L.}~\bibnamefont {Cincio}}, \ and\ \bibinfo {author} {\bibfnamefont {P.~J.}\
  \bibnamefont {Coles}},\ }\bibfield  {title} {\emph {\bibinfo {title}
  {Variational quantum linear solver: A hybrid algorithm for linear systems},\
  }}\href {https://arxiv.org/abs/1909.05820} {\bibfield  {journal} {\bibinfo
  {journal} {arXiv:1909.05820}\ } (\bibinfo {year} {2019})}\BibitemShut
  {NoStop}%
\bibitem [{\citenamefont {Xu}\ \emph {et~al.}(2019)\citenamefont {Xu},
  \citenamefont {Sun}, \citenamefont {Endo}, \citenamefont {Li}, \citenamefont
  {Benjamin},\ and\ \citenamefont {Yuan}}]{xu2019variational}%
  \BibitemOpen
  \bibfield  {author} {\bibinfo {author} {\bibfnamefont {X.}~\bibnamefont
  {Xu}}, \bibinfo {author} {\bibfnamefont {J.}~\bibnamefont {Sun}}, \bibinfo
  {author} {\bibfnamefont {S.}~\bibnamefont {Endo}}, \bibinfo {author}
  {\bibfnamefont {Y.}~\bibnamefont {Li}}, \bibinfo {author} {\bibfnamefont
  {S.~C.}\ \bibnamefont {Benjamin}}, \ and\ \bibinfo {author} {\bibfnamefont
  {X.}~\bibnamefont {Yuan}},\ }\bibfield  {title} {\emph {\bibinfo {title}
  {Variational algorithms for linear algebra},\ }}\href
  {https://arxiv.org/abs/1909.03898} {\bibfield  {journal} {\bibinfo  {journal}
  {arXiv:1909.03898}\ } (\bibinfo {year} {2019})}\BibitemShut {NoStop}%
\bibitem [{\citenamefont {McArdle}\ \emph {et~al.}(2019)\citenamefont
  {McArdle}, \citenamefont {Jones}, \citenamefont {Endo}, \citenamefont {Li},
  \citenamefont {Benjamin},\ and\ \citenamefont
  {Yuan}}]{mcardle2019variational}%
  \BibitemOpen
  \bibfield  {author} {\bibinfo {author} {\bibfnamefont {S.}~\bibnamefont
  {McArdle}}, \bibinfo {author} {\bibfnamefont {T.}~\bibnamefont {Jones}},
  \bibinfo {author} {\bibfnamefont {S.}~\bibnamefont {Endo}}, \bibinfo {author}
  {\bibfnamefont {Y.}~\bibnamefont {Li}}, \bibinfo {author} {\bibfnamefont
  {S.~C.}\ \bibnamefont {Benjamin}}, \ and\ \bibinfo {author} {\bibfnamefont
  {X.}~\bibnamefont {Yuan}},\ }\bibfield  {title} {\emph {\bibinfo {title}
  {Variational ansatz-based quantum simulation of imaginary time evolution},\
  }}\href {\doibasemod 10.1038/s41534-019-0187-2} {\bibfield  {journal}
  {\bibinfo  {journal} {npj Quantum Information}\ }\textbf {\bibinfo {volume}
  {5}},\ \bibinfo {pages} {1} (\bibinfo {year} {2019})}\BibitemShut {NoStop}%
\bibitem [{\citenamefont {Cirstoiu}\ \emph {et~al.}(2019)\citenamefont
  {Cirstoiu}, \citenamefont {Holmes}, \citenamefont {Iosue}, \citenamefont
  {Cincio}, \citenamefont {Coles},\ and\ \citenamefont
  {Sornborger}}]{cirstoiu2019variational}%
  \BibitemOpen
  \bibfield  {author} {\bibinfo {author} {\bibfnamefont {C.}~\bibnamefont
  {Cirstoiu}}, \bibinfo {author} {\bibfnamefont {Z.}~\bibnamefont {Holmes}},
  \bibinfo {author} {\bibfnamefont {J.}~\bibnamefont {Iosue}}, \bibinfo
  {author} {\bibfnamefont {L.}~\bibnamefont {Cincio}}, \bibinfo {author}
  {\bibfnamefont {P.~J.}\ \bibnamefont {Coles}}, \ and\ \bibinfo {author}
  {\bibfnamefont {A.}~\bibnamefont {Sornborger}},\ }\bibfield  {title} {\emph
  {\bibinfo {title} {Variational fast forwarding for quantum simulation beyond
  the coherence time},\ }}\href {https://arxiv.org/abs/1910.04292} {\bibfield
  {journal} {\bibinfo  {journal} {arXiv:1910.04292}\ } (\bibinfo {year}
  {2019})}\BibitemShut {NoStop}%
\bibitem [{\citenamefont {Wecker}\ \emph {et~al.}(2015)\citenamefont {Wecker},
  \citenamefont {Hastings},\ and\ \citenamefont {Troyer}}]{troyer2015}%
  \BibitemOpen
  \bibfield  {author} {\bibinfo {author} {\bibfnamefont {D.}~\bibnamefont
  {Wecker}}, \bibinfo {author} {\bibfnamefont {M.~B.}\ \bibnamefont
  {Hastings}}, \ and\ \bibinfo {author} {\bibfnamefont {M.}~\bibnamefont
  {Troyer}},\ }\bibfield  {title} {\emph {\bibinfo {title} {Progress towards
  practical quantum variational algorithms},\ }}\href {\doibasemod
  10.1103/PhysRevA.92.042303} {\bibfield  {journal} {\bibinfo  {journal} {Phys.
  Rev. A}\ }\textbf {\bibinfo {volume} {92}},\ \bibinfo {pages} {042303}
  (\bibinfo {year} {2015})}\BibitemShut {NoStop}%
\bibitem [{\citenamefont {Cao}\ \emph {et~al.}(2018)\citenamefont {Cao},
  \citenamefont {Romero}, \citenamefont {Olson}, \citenamefont {Degroote},
  \citenamefont {Johnson}, \citenamefont {Kieferov{\'a}}, \citenamefont
  {Kivlichan}, \citenamefont {Menke}, \citenamefont {Peropadre}, \citenamefont
  {Sawaya} \emph {et~al.}}]{cao2018quantum}%
  \BibitemOpen
  \bibfield  {author} {\bibinfo {author} {\bibfnamefont {Y.}~\bibnamefont
  {Cao}}, \bibinfo {author} {\bibfnamefont {J.}~\bibnamefont {Romero}},
  \bibinfo {author} {\bibfnamefont {J.~P.}\ \bibnamefont {Olson}}, \bibinfo
  {author} {\bibfnamefont {M.}~\bibnamefont {Degroote}}, \bibinfo {author}
  {\bibfnamefont {P.~D.}\ \bibnamefont {Johnson}}, \bibinfo {author}
  {\bibfnamefont {M.}~\bibnamefont {Kieferov{\'a}}}, \bibinfo {author}
  {\bibfnamefont {I.~D.}\ \bibnamefont {Kivlichan}}, \bibinfo {author}
  {\bibfnamefont {T.}~\bibnamefont {Menke}}, \bibinfo {author} {\bibfnamefont
  {B.}~\bibnamefont {Peropadre}}, \bibinfo {author} {\bibfnamefont {N.~P.}\
  \bibnamefont {Sawaya}},  \emph {et~al.},\ }\bibfield  {title} {\emph
  {\bibinfo {title} {Quantum chemistry in the age of quantum computing},\
  }}\href {\doibasemod 10.1021/acs.chemrev.8b00803} {\bibfield  {journal}
  {\bibinfo  {journal} {Chemical reviews}\ } (\bibinfo {year} {2018}),\
  10.1021/acs.chemrev.8b00803}\BibitemShut {NoStop}%
\bibitem [{\citenamefont {McArdle}\ \emph {et~al.}(2018)\citenamefont
  {McArdle}, \citenamefont {Endo}, \citenamefont {Aspuru-Guzik}, \citenamefont
  {Benjamin},\ and\ \citenamefont {Yuan}}]{mcardle2018quantum}%
  \BibitemOpen
  \bibfield  {author} {\bibinfo {author} {\bibfnamefont {S.}~\bibnamefont
  {McArdle}}, \bibinfo {author} {\bibfnamefont {S.}~\bibnamefont {Endo}},
  \bibinfo {author} {\bibfnamefont {A.}~\bibnamefont {Aspuru-Guzik}}, \bibinfo
  {author} {\bibfnamefont {S.}~\bibnamefont {Benjamin}}, \ and\ \bibinfo
  {author} {\bibfnamefont {X.}~\bibnamefont {Yuan}},\ }\bibfield  {title}
  {\emph {\bibinfo {title} {Quantum computational chemistry},\ }}\href
  {https://arxiv.org/abs/1808.10402} {\bibfield  {journal} {\bibinfo  {journal}
  {arXiv:1808.10402}\ } (\bibinfo {year} {2018})}\BibitemShut {NoStop}%
\bibitem [{\citenamefont {Jena}\ \emph {et~al.}(2019)\citenamefont {Jena},
  \citenamefont {Genin},\ and\ \citenamefont {Mosca}}]{Jena2019}%
  \BibitemOpen
  \bibfield  {author} {\bibinfo {author} {\bibfnamefont {A.}~\bibnamefont
  {Jena}}, \bibinfo {author} {\bibfnamefont {S.}~\bibnamefont {Genin}}, \ and\
  \bibinfo {author} {\bibfnamefont {M.}~\bibnamefont {Mosca}},\ }\bibfield
  {title} {\emph {\bibinfo {title} {Pauli partitioning with respect to gate
  sets},\ }}\href {https://arxiv.org/abs/1907.07859} {\bibfield  {journal}
  {\bibinfo  {journal} {arXiv:1907.07859}\ } (\bibinfo {year}
  {2019})}\BibitemShut {NoStop}%
\bibitem [{\citenamefont {Izmaylov}\ \emph {et~al.}(2019)\citenamefont
  {Izmaylov}, \citenamefont {Yen}, \citenamefont {Lang},\ and\ \citenamefont
  {Verteletskyi}}]{Izmaylov2019}%
  \BibitemOpen
  \bibfield  {author} {\bibinfo {author} {\bibfnamefont {A.~F.}\ \bibnamefont
  {Izmaylov}}, \bibinfo {author} {\bibfnamefont {T.-C.}\ \bibnamefont {Yen}},
  \bibinfo {author} {\bibfnamefont {R.~A.}\ \bibnamefont {Lang}}, \ and\
  \bibinfo {author} {\bibfnamefont {V.}~\bibnamefont {Verteletskyi}},\
  }\bibfield  {title} {\emph {\bibinfo {title} {Unitary partitioning approach
  to the measurement problem in the variational quantum eigensolver method},\
  }}\href {https://arxiv.org/abs/1907.09040} {\bibfield  {journal} {\bibinfo
  {journal} {arXiv:1907.09040}\ } (\bibinfo {year} {2019})}\BibitemShut
  {NoStop}%
\bibitem [{\citenamefont {Yen}\ \emph {et~al.}(2019)\citenamefont {Yen},
  \citenamefont {Verteletskyi},\ and\ \citenamefont {Izmaylov}}]{Yen2019}%
  \BibitemOpen
  \bibfield  {author} {\bibinfo {author} {\bibfnamefont {T.-C.}\ \bibnamefont
  {Yen}}, \bibinfo {author} {\bibfnamefont {V.}~\bibnamefont {Verteletskyi}}, \
  and\ \bibinfo {author} {\bibfnamefont {A.~F.}\ \bibnamefont {Izmaylov}},\
  }\bibfield  {title} {\emph {\bibinfo {title} {Measuring all compatible
  operators in one series of a single-qubit measurements using unitary
  transformations},\ }}\href {https://arxiv.org/abs/1907.09386} {\bibfield
  {journal} {\bibinfo  {journal} {arXiv:1907.09386}\ } (\bibinfo {year}
  {2019})}\BibitemShut {NoStop}%
\bibitem [{\citenamefont {Gokhale}\ \emph {et~al.}(2019)\citenamefont
  {Gokhale}, \citenamefont {Angiuli}, \citenamefont {Ding}, \citenamefont
  {Gui}, \citenamefont {Tomesh}, \citenamefont {Suchara}, \citenamefont
  {Martonosi},\ and\ \citenamefont {Chong}}]{Gokhale2019}%
  \BibitemOpen
  \bibfield  {author} {\bibinfo {author} {\bibfnamefont {P.}~\bibnamefont
  {Gokhale}}, \bibinfo {author} {\bibfnamefont {O.}~\bibnamefont {Angiuli}},
  \bibinfo {author} {\bibfnamefont {Y.}~\bibnamefont {Ding}}, \bibinfo {author}
  {\bibfnamefont {K.}~\bibnamefont {Gui}}, \bibinfo {author} {\bibfnamefont
  {T.}~\bibnamefont {Tomesh}}, \bibinfo {author} {\bibfnamefont
  {M.}~\bibnamefont {Suchara}}, \bibinfo {author} {\bibfnamefont
  {M.}~\bibnamefont {Martonosi}}, \ and\ \bibinfo {author} {\bibfnamefont
  {F.~T.}\ \bibnamefont {Chong}},\ }\bibfield  {title} {\emph {\bibinfo {title}
  {Minimizing state preparations in variational quantum eigensolver by
  partitioning into commuting families},\ }}\href
  {https://arxiv.org/abs/1907.13623} {\bibfield  {journal} {\bibinfo  {journal}
  {arXiv:1907.13623}\ } (\bibinfo {year} {2019})}\BibitemShut {NoStop}%
\bibitem [{\citenamefont {Crawford}\ \emph {et~al.}(2019)\citenamefont
  {Crawford}, \citenamefont {van Straaten}, \citenamefont {Wang}, \citenamefont
  {Parks}, \citenamefont {Campbell},\ and\ \citenamefont
  {Brierley}}]{Crawford2019}%
  \BibitemOpen
  \bibfield  {author} {\bibinfo {author} {\bibfnamefont {O.}~\bibnamefont
  {Crawford}}, \bibinfo {author} {\bibfnamefont {B.}~\bibnamefont {van
  Straaten}}, \bibinfo {author} {\bibfnamefont {D.}~\bibnamefont {Wang}},
  \bibinfo {author} {\bibfnamefont {T.}~\bibnamefont {Parks}}, \bibinfo
  {author} {\bibfnamefont {E.}~\bibnamefont {Campbell}}, \ and\ \bibinfo
  {author} {\bibfnamefont {S.}~\bibnamefont {Brierley}},\ }\bibfield  {title}
  {\emph {\bibinfo {title} {Efficient quantum measurement of pauli operators},\
  }}\href {https://arxiv.org/abs/1908.06942} {\bibfield  {journal} {\bibinfo
  {journal} {arXiv:1908.06942}\ } (\bibinfo {year} {2019})}\BibitemShut
  {NoStop}%
\bibitem [{\citenamefont {Gokhale}\ and\ \citenamefont
  {Chong}(2019)}]{Gokhale2019-2}%
  \BibitemOpen
  \bibfield  {author} {\bibinfo {author} {\bibfnamefont {P.}~\bibnamefont
  {Gokhale}}\ and\ \bibinfo {author} {\bibfnamefont {F.~T.}\ \bibnamefont
  {Chong}},\ }\bibfield  {title} {\emph {\bibinfo {title} {$o(n^3)$ measurement
  cost for variational quantum eigensolver on molecular hamiltonians},\ }}\href
  {https://arxiv.org/abs/1908.11857} {\bibfield  {journal} {\bibinfo  {journal}
  {arXiv:1908.11857}\ } (\bibinfo {year} {2019})}\BibitemShut {NoStop}%
\bibitem [{\citenamefont {Huggins}\ \emph {et~al.}(2019)\citenamefont
  {Huggins}, \citenamefont {McClean}, \citenamefont {Rubin}, \citenamefont
  {Jiang}, \citenamefont {Wiebe}, \citenamefont {Whaley},\ and\ \citenamefont
  {Babbush}}]{huggins2019efficient}%
  \BibitemOpen
  \bibfield  {author} {\bibinfo {author} {\bibfnamefont {W.~J.}\ \bibnamefont
  {Huggins}}, \bibinfo {author} {\bibfnamefont {J.}~\bibnamefont {McClean}},
  \bibinfo {author} {\bibfnamefont {N.}~\bibnamefont {Rubin}}, \bibinfo
  {author} {\bibfnamefont {Z.}~\bibnamefont {Jiang}}, \bibinfo {author}
  {\bibfnamefont {N.}~\bibnamefont {Wiebe}}, \bibinfo {author} {\bibfnamefont
  {K.~B.}\ \bibnamefont {Whaley}}, \ and\ \bibinfo {author} {\bibfnamefont
  {R.}~\bibnamefont {Babbush}},\ }\bibfield  {title} {\emph {\bibinfo {title}
  {Efficient and noise resilient measurements for quantum chemistry on
  near-term quantum computers},\ }}\href {https://arxiv.org/abs/1907.13117}
  {\bibfield  {journal} {\bibinfo  {journal} {arXiv:1907.13117}\ } (\bibinfo
  {year} {2019})}\BibitemShut {NoStop}%
\bibitem [{\citenamefont {Verdon}\ \emph {et~al.}(2019)\citenamefont {Verdon},
  \citenamefont {Broughton}, \citenamefont {McClean}, \citenamefont {Sung},
  \citenamefont {Babbush}, \citenamefont {Jiang}, \citenamefont {Neven},\ and\
  \citenamefont {Mohseni}}]{verdon2019learning}%
  \BibitemOpen
  \bibfield  {author} {\bibinfo {author} {\bibfnamefont {G.}~\bibnamefont
  {Verdon}}, \bibinfo {author} {\bibfnamefont {M.}~\bibnamefont {Broughton}},
  \bibinfo {author} {\bibfnamefont {J.~R.}\ \bibnamefont {McClean}}, \bibinfo
  {author} {\bibfnamefont {K.~J.}\ \bibnamefont {Sung}}, \bibinfo {author}
  {\bibfnamefont {R.}~\bibnamefont {Babbush}}, \bibinfo {author} {\bibfnamefont
  {Z.}~\bibnamefont {Jiang}}, \bibinfo {author} {\bibfnamefont
  {H.}~\bibnamefont {Neven}}, \ and\ \bibinfo {author} {\bibfnamefont
  {M.}~\bibnamefont {Mohseni}},\ }\bibfield  {title} {\emph {\bibinfo {title}
  {Learning to learn with quantum neural networks via classical neural
  networks},\ }}\href {https://arxiv.org/abs/1907.05415} {\bibfield  {journal}
  {\bibinfo  {journal} {arXiv:1907.05415}\ } (\bibinfo {year}
  {2019})}\BibitemShut {NoStop}%
\bibitem [{\citenamefont {Wilson}\ \emph {et~al.}(2019)\citenamefont {Wilson},
  \citenamefont {Stromswold}, \citenamefont {Wudarski}, \citenamefont
  {Hadfield}, \citenamefont {Tubman},\ and\ \citenamefont
  {Rieffel}}]{wilson2019optimizing}%
  \BibitemOpen
  \bibfield  {author} {\bibinfo {author} {\bibfnamefont {M.}~\bibnamefont
  {Wilson}}, \bibinfo {author} {\bibfnamefont {S.}~\bibnamefont {Stromswold}},
  \bibinfo {author} {\bibfnamefont {F.}~\bibnamefont {Wudarski}}, \bibinfo
  {author} {\bibfnamefont {S.}~\bibnamefont {Hadfield}}, \bibinfo {author}
  {\bibfnamefont {N.~M.}\ \bibnamefont {Tubman}}, \ and\ \bibinfo {author}
  {\bibfnamefont {E.}~\bibnamefont {Rieffel}},\ }\bibfield  {title} {\emph
  {\bibinfo {title} {Optimizing quantum heuristics with meta-learning},\
  }}\href {https://arxiv.org/abs/1908.03185} {\bibfield  {journal} {\bibinfo
  {journal} {arXiv:1908.03185}\ } (\bibinfo {year} {2019})}\BibitemShut
  {NoStop}%
\bibitem [{\citenamefont {Nakanishi}\ \emph {et~al.}(2019)\citenamefont
  {Nakanishi}, \citenamefont {Fujii},\ and\ \citenamefont
  {Todo}}]{nakanishi2019}%
  \BibitemOpen
  \bibfield  {author} {\bibinfo {author} {\bibfnamefont {K.~M.}\ \bibnamefont
  {Nakanishi}}, \bibinfo {author} {\bibfnamefont {K.}~\bibnamefont {Fujii}}, \
  and\ \bibinfo {author} {\bibfnamefont {S.}~\bibnamefont {Todo}},\ }\bibfield
  {title} {\emph {\bibinfo {title} {Sequential minimal optimization for
  quantum-classical hybrid algorithms},\ }}\href
  {https://arxiv.org/abs/1903.12166} {\bibfield  {journal} {\bibinfo  {journal}
  {arXiv:1903.12166}\ } (\bibinfo {year} {2019})}\BibitemShut {NoStop}%
\bibitem [{\citenamefont {Parrish}\ \emph {et~al.}(2019)\citenamefont
  {Parrish}, \citenamefont {Iosue}, \citenamefont {Ozaeta},\ and\ \citenamefont
  {McMahon}}]{parrish2019}%
  \BibitemOpen
  \bibfield  {author} {\bibinfo {author} {\bibfnamefont {R.~M.}\ \bibnamefont
  {Parrish}}, \bibinfo {author} {\bibfnamefont {J.~T.}\ \bibnamefont {Iosue}},
  \bibinfo {author} {\bibfnamefont {A.}~\bibnamefont {Ozaeta}}, \ and\ \bibinfo
  {author} {\bibfnamefont {P.~L.}\ \bibnamefont {McMahon}},\ }\bibfield
  {title} {\emph {\bibinfo {title} {A {J}acobi diagonalization and {A}nderson
  acceleration algorithm for variational quantum algorithm parameter
  optimization},\ }}\href {https://arxiv.org/abs/1904.03206} {\bibfield
  {journal} {\bibinfo  {journal} {arXiv:1904.03206}\ } (\bibinfo {year}
  {2019})}\BibitemShut {NoStop}%
\bibitem [{\citenamefont {Stokes}\ \emph {et~al.}(2019)\citenamefont {Stokes},
  \citenamefont {Izaac}, \citenamefont {Killoran},\ and\ \citenamefont
  {Carleo}}]{stokes2019quantum}%
  \BibitemOpen
  \bibfield  {author} {\bibinfo {author} {\bibfnamefont {J.}~\bibnamefont
  {Stokes}}, \bibinfo {author} {\bibfnamefont {J.}~\bibnamefont {Izaac}},
  \bibinfo {author} {\bibfnamefont {N.}~\bibnamefont {Killoran}}, \ and\
  \bibinfo {author} {\bibfnamefont {G.}~\bibnamefont {Carleo}},\ }\bibfield
  {title} {\emph {\bibinfo {title} {Quantum natural gradient},\ }}\href
  {https://arxiv.org/abs/1909.02108} {\bibfield  {journal} {\bibinfo  {journal}
  {arXiv:1909.02108}\ } (\bibinfo {year} {2019})}\BibitemShut {NoStop}%
\bibitem [{\citenamefont {Balles}\ \emph {et~al.}(2017)\citenamefont {Balles},
  \citenamefont {Romero},\ and\ \citenamefont {Hennig}}]{Balles2017}%
  \BibitemOpen
  \bibfield  {author} {\bibinfo {author} {\bibfnamefont {L.}~\bibnamefont
  {Balles}}, \bibinfo {author} {\bibfnamefont {J.}~\bibnamefont {Romero}}, \
  and\ \bibinfo {author} {\bibfnamefont {P.}~\bibnamefont {Hennig}},\ }in\
  \href {http://auai.org/uai2017/proceedings/papers/141.pdf} {\emph {\bibinfo
  {booktitle} {Proceedings of the Thirty-Third Conference on Uncertainty in
  Artificial Intelligence (UAI)}}}\ (\bibinfo {year} {2017})\ pp.\ \bibinfo
  {pages} {410--419}\BibitemShut {NoStop}%
\bibitem [{\citenamefont {et.al.}(2019)}]{gadi_aleksandrowicz_2019_2562111}%
  \BibitemOpen
  \bibfield  {author} {\bibinfo {author} {\bibfnamefont {G.~A.}\ \bibnamefont
  {et.al.}},\ }\href {\doibasemod 10.5281/zenodo.2562111} {\bibinfo {title}
  {{Qiskit: An Open-source Framework for Quantum Computing}},\ } (\bibinfo
  {year} {2019})\BibitemShut {NoStop}%
\bibitem [{\citenamefont {Kingma}\ and\ \citenamefont {Ba}(2015)}]{Kingma2015}%
  \BibitemOpen
  \bibfield  {author} {\bibinfo {author} {\bibfnamefont {D.~P.}\ \bibnamefont
  {Kingma}}\ and\ \bibinfo {author} {\bibfnamefont {J.}~\bibnamefont {Ba}},\
  }in\ \href {http://arxiv.org/abs/1412.6980} {\emph {\bibinfo {booktitle}
  {Proceedings of the 3rd International Conference on Learning Representations
  (ICLR)}}}\ (\bibinfo {year} {2015})\BibitemShut {NoStop}%
\bibitem [{\citenamefont {Spall}(1992)}]{spall1992}%
  \BibitemOpen
  \bibfield  {author} {\bibinfo {author} {\bibfnamefont {J.~C.}\ \bibnamefont
  {Spall}},\ }\bibfield  {title} {\emph {\bibinfo {title} {Multivariate
  stochastic approximation using a simultaneous perturbation gradient
  approximation},\ }}\href {\doibasemod 10.1109/9.119632} {\bibfield  {journal}
  {\bibinfo  {journal} {IEEE transactions on automatic control}\ }\textbf
  {\bibinfo {volume} {37}},\ \bibinfo {pages} {332} (\bibinfo {year}
  {1992})}\BibitemShut {NoStop}%
\bibitem [{\citenamefont {LeCun}\ \emph {et~al.}(2015)\citenamefont {LeCun},
  \citenamefont {Bengio},\ and\ \citenamefont {Hinton}}]{lecun2015}%
  \BibitemOpen
  \bibfield  {author} {\bibinfo {author} {\bibfnamefont {Y.}~\bibnamefont
  {LeCun}}, \bibinfo {author} {\bibfnamefont {Y.}~\bibnamefont {Bengio}}, \
  and\ \bibinfo {author} {\bibfnamefont {G.}~\bibnamefont {Hinton}},\
  }\bibfield  {title} {\emph {\bibinfo {title} {Deep learning},\ }}\href
  {\doibasemod 10.1038/nature14539} {\bibfield  {journal} {\bibinfo  {journal}
  {Nature}\ }\textbf {\bibinfo {volume} {521}},\ \bibinfo {pages} {436}
  (\bibinfo {year} {2015})}\BibitemShut {NoStop}%
\bibitem [{\citenamefont {Mitarai}\ \emph {et~al.}(2018)\citenamefont
  {Mitarai}, \citenamefont {Negoro}, \citenamefont {Kitagawa},\ and\
  \citenamefont {Fujii}}]{Mitarai2018}%
  \BibitemOpen
  \bibfield  {author} {\bibinfo {author} {\bibfnamefont {K.}~\bibnamefont
  {Mitarai}}, \bibinfo {author} {\bibfnamefont {M.}~\bibnamefont {Negoro}},
  \bibinfo {author} {\bibfnamefont {M.}~\bibnamefont {Kitagawa}}, \ and\
  \bibinfo {author} {\bibfnamefont {K.}~\bibnamefont {Fujii}},\ }\bibfield
  {title} {\emph {\bibinfo {title} {Quantum circuit learning},\ }}\href
  {\doibasemod 10.1103/PhysRevA.98.032309} {\bibfield  {journal} {\bibinfo
  {journal} {Phys. Rev. A}\ }\textbf {\bibinfo {volume} {98}},\ \bibinfo
  {pages} {032309} (\bibinfo {year} {2018})}\BibitemShut {NoStop}%
\bibitem [{\citenamefont {Schuld}\ \emph {et~al.}(2019)\citenamefont {Schuld},
  \citenamefont {Bergholm}, \citenamefont {Gogolin}, \citenamefont {Izaac},\
  and\ \citenamefont {Killoran}}]{Schuld2019}%
  \BibitemOpen
  \bibfield  {author} {\bibinfo {author} {\bibfnamefont {M.}~\bibnamefont
  {Schuld}}, \bibinfo {author} {\bibfnamefont {V.}~\bibnamefont {Bergholm}},
  \bibinfo {author} {\bibfnamefont {C.}~\bibnamefont {Gogolin}}, \bibinfo
  {author} {\bibfnamefont {J.}~\bibnamefont {Izaac}}, \ and\ \bibinfo {author}
  {\bibfnamefont {N.}~\bibnamefont {Killoran}},\ }\bibfield  {title} {\emph
  {\bibinfo {title} {Evaluating analytic gradients on quantum hardware},\
  }}\href {\doibasemod 10.1103/PhysRevA.99.032331} {\bibfield  {journal}
  {\bibinfo  {journal} {Phys. Rev. A}\ }\textbf {\bibinfo {volume} {99}},\
  \bibinfo {pages} {032331} (\bibinfo {year} {2019})}\BibitemShut {NoStop}%
\bibitem [{\citenamefont {Bergholm}\ \emph {et~al.}(2018)\citenamefont
  {Bergholm}, \citenamefont {Izaac}, \citenamefont {Schuld}, \citenamefont
  {Gogolin},\ and\ \citenamefont {Killoran}}]{bergholm2018pennylane}%
  \BibitemOpen
  \bibfield  {author} {\bibinfo {author} {\bibfnamefont {V.}~\bibnamefont
  {Bergholm}}, \bibinfo {author} {\bibfnamefont {J.}~\bibnamefont {Izaac}},
  \bibinfo {author} {\bibfnamefont {M.}~\bibnamefont {Schuld}}, \bibinfo
  {author} {\bibfnamefont {C.}~\bibnamefont {Gogolin}}, \ and\ \bibinfo
  {author} {\bibfnamefont {N.}~\bibnamefont {Killoran}},\ }\bibfield  {title}
  {\emph {\bibinfo {title} {Pennylane: Automatic differentiation of hybrid
  quantum-classical computations},\ }}\href {https://arxiv.org/abs/1811.04968}
  {\bibfield  {journal} {\bibinfo  {journal} {arXiv:1811.04968}\ } (\bibinfo
  {year} {2018})}\BibitemShut {NoStop}%
\bibitem [{\citenamefont {Harrow}\ and\ \citenamefont
  {Napp}(2019)}]{harrow2019}%
  \BibitemOpen
  \bibfield  {author} {\bibinfo {author} {\bibfnamefont {A.}~\bibnamefont
  {Harrow}}\ and\ \bibinfo {author} {\bibfnamefont {J.}~\bibnamefont {Napp}},\
  }\bibfield  {title} {\emph {\bibinfo {title} {Low-depth gradient measurements
  can improve convergence in variational hybrid quantum-classical algorithms},\
  }}\href {https://arxiv.org/abs/1901.05374} {\bibfield  {journal} {\bibinfo
  {journal} {arXiv:1901.05374}\ } (\bibinfo {year} {2019})}\BibitemShut
  {NoStop}%
\bibitem [{\citenamefont {Guerreschi}\ and\ \citenamefont
  {Smelyanskiy}(2017)}]{Guerreschi2017}%
  \BibitemOpen
  \bibfield  {author} {\bibinfo {author} {\bibfnamefont {G.~G.}\ \bibnamefont
  {Guerreschi}}\ and\ \bibinfo {author} {\bibfnamefont {M.}~\bibnamefont
  {Smelyanskiy}},\ }\bibfield  {title} {\emph {\bibinfo {title} {Practical
  optimization for hybrid quantum-classical algorithms},\ }}\href
  {https://arxiv.org/abs/1701.01450} {\bibfield  {journal} {\bibinfo  {journal}
  {arXiv:1701.01450}\ } (\bibinfo {year} {2017})}\BibitemShut {NoStop}%
\bibitem [{\citenamefont {Bergstra}\ \emph {et~al.}(2011)\citenamefont
  {Bergstra}, \citenamefont {Bardenet}, \citenamefont {Bengio},\ and\
  \citenamefont {K{\'e}gl}}]{Bergstra2011}%
  \BibitemOpen
  \bibfield  {author} {\bibinfo {author} {\bibfnamefont {J.~S.}\ \bibnamefont
  {Bergstra}}, \bibinfo {author} {\bibfnamefont {R.}~\bibnamefont {Bardenet}},
  \bibinfo {author} {\bibfnamefont {Y.}~\bibnamefont {Bengio}}, \ and\ \bibinfo
  {author} {\bibfnamefont {B.}~\bibnamefont {K{\'e}gl}},\ }in\ \href
  {http://papers.nips.cc/paper/4443-algorithms-for-hyper-parameter-optimization.pdf}
  {\emph {\bibinfo {booktitle} {Advances in Neural Information Processing
  Systems 24}}}\ (\bibinfo {year} {2011})\ pp.\ \bibinfo {pages}
  {2546--2554}\BibitemShut {NoStop}%
\bibitem [{\citenamefont {Liu}\ \emph {et~al.}(2019)\citenamefont {Liu},
  \citenamefont {Jiang}, \citenamefont {He}, \citenamefont {Chen},
  \citenamefont {Liu}, \citenamefont {Gao},\ and\ \citenamefont
  {Han}}]{liu2019variance}%
  \BibitemOpen
  \bibfield  {author} {\bibinfo {author} {\bibfnamefont {L.}~\bibnamefont
  {Liu}}, \bibinfo {author} {\bibfnamefont {H.}~\bibnamefont {Jiang}}, \bibinfo
  {author} {\bibfnamefont {P.}~\bibnamefont {He}}, \bibinfo {author}
  {\bibfnamefont {W.}~\bibnamefont {Chen}}, \bibinfo {author} {\bibfnamefont
  {X.}~\bibnamefont {Liu}}, \bibinfo {author} {\bibfnamefont {J.}~\bibnamefont
  {Gao}}, \ and\ \bibinfo {author} {\bibfnamefont {J.}~\bibnamefont {Han}},\
  }\bibfield  {title} {\emph {\bibinfo {title} {On the variance of the adaptive
  learning rate and beyond},\ }}\href {https://arxiv.org/abs/1908.03265}
  {\bibfield  {journal} {\bibinfo  {journal} {arXiv:1908.03265}\ } (\bibinfo
  {year} {2019})}\BibitemShut {NoStop}%
\bibitem [{\citenamefont {Spall}(1998)}]{spall1998}%
  \BibitemOpen
  \bibfield  {author} {\bibinfo {author} {\bibfnamefont {J.~C.}\ \bibnamefont
  {Spall}},\ }\bibfield  {title} {\emph {\bibinfo {title} {Implementation of
  the simultaneous perturbation algorithm for stochastic optimization},\
  }}\href {\doibasemod 10.1109/7.705889} {\bibfield  {journal} {\bibinfo
  {journal} {IEEE Transactions on aerospace and electronic systems}\ }\textbf
  {\bibinfo {volume} {34}},\ \bibinfo {pages} {817} (\bibinfo {year}
  {1998})}\BibitemShut {NoStop}%
\bibitem [{\citenamefont {Kandala}\ \emph {et~al.}(2017)\citenamefont
  {Kandala}, \citenamefont {Mezzacapo}, \citenamefont {Temme}, \citenamefont
  {Takita}, \citenamefont {Brink}, \citenamefont {Chow},\ and\ \citenamefont
  {Gambetta}}]{kandala2017}%
  \BibitemOpen
  \bibfield  {author} {\bibinfo {author} {\bibfnamefont {A.}~\bibnamefont
  {Kandala}}, \bibinfo {author} {\bibfnamefont {A.}~\bibnamefont {Mezzacapo}},
  \bibinfo {author} {\bibfnamefont {K.}~\bibnamefont {Temme}}, \bibinfo
  {author} {\bibfnamefont {M.}~\bibnamefont {Takita}}, \bibinfo {author}
  {\bibfnamefont {M.}~\bibnamefont {Brink}}, \bibinfo {author} {\bibfnamefont
  {J.~M.}\ \bibnamefont {Chow}}, \ and\ \bibinfo {author} {\bibfnamefont
  {J.~M.}\ \bibnamefont {Gambetta}},\ }\bibfield  {title} {\emph {\bibinfo
  {title} {Hardware-efficient variational quantum eigensolver for small
  molecules and quantum magnets},\ }}\href {\doibasemod 10.1038/nature23879}
  {\bibfield  {journal} {\bibinfo  {journal} {Nature}\ }\textbf {\bibinfo
  {volume} {549}},\ \bibinfo {pages} {242} (\bibinfo {year}
  {2017})}\BibitemShut {NoStop}%
\bibitem [{\citenamefont {Powell}(1964)}]{powell1964}%
  \BibitemOpen
  \bibfield  {author} {\bibinfo {author} {\bibfnamefont {M.~J.}\ \bibnamefont
  {Powell}},\ }\bibfield  {title} {\emph {\bibinfo {title} {{An efficient
  method for finding the minimum of a function of several variables without
  calculating derivatives}},\ }}\href {\doibasemod 10.1093/comjnl/7.2.155}
  {\bibfield  {journal} {\bibinfo  {journal} {The Computer Journal}\ }\textbf
  {\bibinfo {volume} {7}},\ \bibinfo {pages} {155} (\bibinfo {year}
  {1964})}\BibitemShut {NoStop}%
\bibitem [{\citenamefont {Brent}(2013)}]{Brent2013}%
  \BibitemOpen
  \bibfield  {author} {\bibinfo {author} {\bibfnamefont {R.~P.}\ \bibnamefont
  {Brent}},\ }\href@noop {} {\emph {\bibinfo {title} {Algorithms for
  Minimization Without Derivatives}}}\ (\bibinfo  {publisher} {Dover
  Publications},\ \bibinfo {year} {2013})\BibitemShut {NoStop}%
\bibitem [{\citenamefont {Anderson}(1965)}]{anderson1965iterative}%
  \BibitemOpen
  \bibfield  {author} {\bibinfo {author} {\bibfnamefont {D.~G.}\ \bibnamefont
  {Anderson}},\ }\bibfield  {title} {\emph {\bibinfo {title} {Iterative
  procedures for nonlinear integral equations},\ }}\href {\doibasemod
  10.1145/321296.321305} {\bibfield  {journal} {\bibinfo  {journal} {Journal of
  the ACM (JACM)}\ }\textbf {\bibinfo {volume} {12}},\ \bibinfo {pages} {547}
  (\bibinfo {year} {1965})}\BibitemShut {NoStop}%
\bibitem [{\citenamefont {Pulay}(1982)}]{pulay1982improved}%
  \BibitemOpen
  \bibfield  {author} {\bibinfo {author} {\bibfnamefont {P.}~\bibnamefont
  {Pulay}},\ }\bibfield  {title} {\emph {\bibinfo {title} {Improved scf
  convergence acceleration},\ }}\href {\doibasemod 10.1002/jcc.540030413}
  {\bibfield  {journal} {\bibinfo  {journal} {Journal of Computational
  Chemistry}\ }\textbf {\bibinfo {volume} {3}},\ \bibinfo {pages} {556}
  (\bibinfo {year} {1982})}\BibitemShut {NoStop}%
\bibitem [{IBM(2018)}]{IBMQ14}%
  \BibitemOpen
  \href@noop {} {\bibinfo {title} {{IBM} {Q} 16 {M}elbourne backend
  specification},\ }\bibinfo {howpublished}
  {\url{https://github.com/Qiskit/ibmq-device-information/tree/master/backends/melbourne/V1}}
  (\bibinfo {year} {2018})\BibitemShut {NoStop}%
\bibitem [{\citenamefont {Sweke}\ \emph {et~al.}(2019)\citenamefont {Sweke},
  \citenamefont {Wilde}, \citenamefont {Meyer}, \citenamefont {Schuld},
  \citenamefont {F{\"a}hrmann}, \citenamefont {Meynard-Piganeau},\ and\
  \citenamefont {Eisert}}]{sweke2019stochastic}%
  \BibitemOpen
  \bibfield  {author} {\bibinfo {author} {\bibfnamefont {R.}~\bibnamefont
  {Sweke}}, \bibinfo {author} {\bibfnamefont {F.}~\bibnamefont {Wilde}},
  \bibinfo {author} {\bibfnamefont {J.}~\bibnamefont {Meyer}}, \bibinfo
  {author} {\bibfnamefont {M.}~\bibnamefont {Schuld}}, \bibinfo {author}
  {\bibfnamefont {P.~K.}\ \bibnamefont {F{\"a}hrmann}}, \bibinfo {author}
  {\bibfnamefont {B.}~\bibnamefont {Meynard-Piganeau}}, \ and\ \bibinfo
  {author} {\bibfnamefont {J.}~\bibnamefont {Eisert}},\ }\bibfield  {title}
  {\emph {\bibinfo {title} {Stochastic gradient descent for hybrid
  quantum-classical optimization},\ }}\href {https://arxiv.org/abs/1910.01155}
  {\bibfield  {journal} {\bibinfo  {journal} {arXiv:1910.01155}\ } (\bibinfo
  {year} {2019})}\BibitemShut {NoStop}%
\end{thebibliography}
%

\appendix

\section{The Expected Lower Bound on the Gain per Shot}\label{App:Derivation}

Here we repeat the derivation provided by~\cite{Balles2017} for the lower bound on the expected gain per shot (given in \eqref{eq:expected_gain}), and extend it to our expression lower bounding the expected gain per shot per partial derivative \eqref{eq:gain_per_shot}. 

Assuming that the cost function is admits a Taylor series representation about the current point in parameter space, to quadratic order we have
\begin{multline}
f(\fattheta') = f(\fattheta)+ \sum_{i=1}^d (\theta_i'-\theta_i)\partial_i f(\fattheta) \\
+\frac{1}{2} \sum_{i=1}^d \sum_{j=1}^d(\theta_i'-\theta_i)(\theta_j'-\theta_j)\partial_i\partial_j f(\fattheta).
\end{multline}
In this way, we approximate the gain (the change in the cost function) we expect after the update step, with $\fattheta'=\fattheta-\alpha \vec{g}$:
\begin{multline}
f(\fattheta)-f(\fattheta') = \alpha\sum_{i=1}^d g_i\partial_i f(\fattheta) \\
-\frac{\alpha^2}{2} \sum_{i=1}^d \sum_{j=1}^d g_i g_j\partial_i\partial_j f(\fattheta).
\end{multline}
If the gradients are Lipschitz continuous, we can achieve a lower bound  $\mathcal{G}$ on this quantity using the Lipschitz constant $L$:
\begin{equation}\label{eq:gain_bound}
    \mathcal{G} = \alpha\vec{\nabla} f(\fattheta)\cdot\vec{g}-\frac{\alpha^2L}{2} \|\vec{g}\|^2.
\end{equation}

Next we assume that the gradient estimates $\vec{g}$ have mean $\mathbb{E}\left[\vec{g}\right]=\vec{\nabla}f(\fattheta)$ and covariance $\Sigma/s$, where $s$ is the number of shots used in the estimate. We then have 
\begin{multline}
    \mathbb{E}\left[\|\vec{g}\|^2\right]= \sum_{i=1}^d\mathbb{E}\left[g_i^2\right]=\|\vec{\nabla}f(\fattheta)\|^2+\sum_{i=1}^d\Sigma_{ii}/s.
\end{multline}
Plugging this back into \eqref{eq:gain_bound} then gives us 
\begin{equation}\label{eq:expected_gain2}
    \mathbb{E}\left[\mathcal{G}\right] = \left( \alpha - \frac{L \alpha^2}{2}\right)\|\vec{\nabla} f\|^2- \frac{L\alpha^2}{2s}\Tr (\Sigma),
\end{equation}
which is \eqref{eq:expected_gain}. Dividing both sides by $s$ then gives the expected lower bound on the gain per shot. In order to arrive at \eqref{eq:gain_per_shot}, we rewrite this expression as:
\begin{align}\label{eq:expected_gain_sum}
    \mathbb{E}\left[\mathcal{G}\right] &=\sum_{i=1}^d\left[ \left( \alpha - \frac{L \alpha^2}{2}\right)(\partial_i f)^2- \frac{L\alpha^2}{2s_i}\Sigma_{ii}\right]\\ \nonumber
    &=\sum_{i=1}^d \mathbb{E}\left[\mathcal{G}_i\right]
\end{align}
Finally, defining  $\gamma_i=\mathbb{E}\left[\mathcal{G}_i\right]/s_i$ and replacing $\partial_i f$ and $\Sigma_{ii}$ with their estimators $g_i$ and $S_i$, respectively, gives \eqref{eq:gain_per_shot}.

\section{CANS Algorithm}\label{App:CANS}

For the interested reader, we present the algorithm for CANS (Coupled Adaptive Number of Shots) in Algorithm \ref{alg:CANS}, which is an adaptation of the CABS algorithm \cite{Balles2017} to the VHQCA setting.

\begin{figure}[h]
\begin{algorithm}[H]
\begin{algorithmic}[1]
\Statex \textbf{Input:} Learning rate $\alpha$, starting point $\fattheta_0$, min number of shots per estimation  $s_\text{min}$, number of shots that can be used in total $N$, Lipschitz constant $L$, running average constant $\mu$, bias for gradient norm $b$
\State initialize: $\fattheta \leftarrow \fattheta_0 $, $s_\text{tot} \leftarrow 0$, $s \leftarrow s_\text{min}$, $\vec{\chi} \leftarrow (0,...,0)^T$, $\xi \leftarrow 0$, $k\gets0$
\While{$s_{\tot} < N$}
    \State $\vec{g}, \vec{S} \gets Evaluate(\fattheta, s)$
\State $s_{\tot} \gets s_{\tot} + 2 s$
\State $\fattheta \gets \fattheta - \alpha \vec{g}$
\State $\xi \leftarrow \mu \xi + (1-\mu) \|\vec{S}\|_1$
\State  $\vec{\chi} \leftarrow \mu \vec{\chi} + (1-\mu) \vec{g}$
\State$s \leftarrow \left\lceil\frac{2L\alpha}{2-L\alpha} \frac{\xi}{\|\chi\|^2 + b\mu^k}\right\rceil$
\State $\vec{s} \gets \max (s,s_\text{min})$
\State $k \gets k+1$
\EndWhile
\end{algorithmic}
\caption{\justified{Stochastic gradient descent with CANS. The function {$Evaluate(\fattheta, s)$} evaluates the gradient at $\fattheta$ using $s$ measurements for each component of the derivative using the parameter shift rule  \eqref{eq:analytic_gradient} and returns the estimated gradient vector $\vec{g}$ as well as the vector $\vec{S}$ with the variances of the individual estimates of the partial derivatives.}}
\label{alg:CANS}
\end{algorithm}
\end{figure}

\section{Cumulative probability distributions for 3-qubit implementations}\label{App:Cumulative}

\begin{figure*}[!t]
    \centering
    \includegraphics[width=1.8\columnwidth]{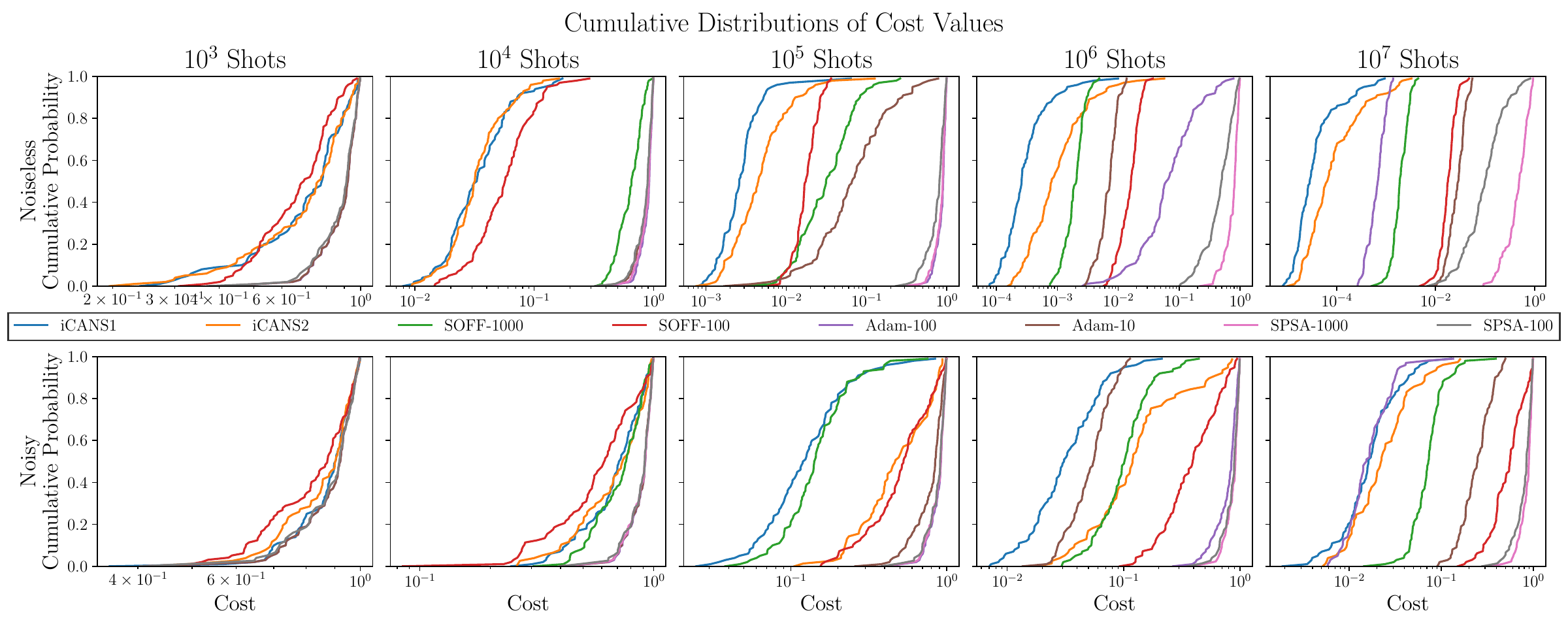}%
    \caption{Comparison of performance for the varaitional compiling task across one hundred random starts. Each panel shows the cumulative probability distribution of the cost values acheived by each optimizer for a different value of total shots $N$. Note that the further to the left a curve is, the better the optimizer has minimized the cost. }
    \label{fig:compilation_cumulative}
\end{figure*}

\begin{figure*}
    \centering
    \includegraphics[width=1.8\columnwidth]{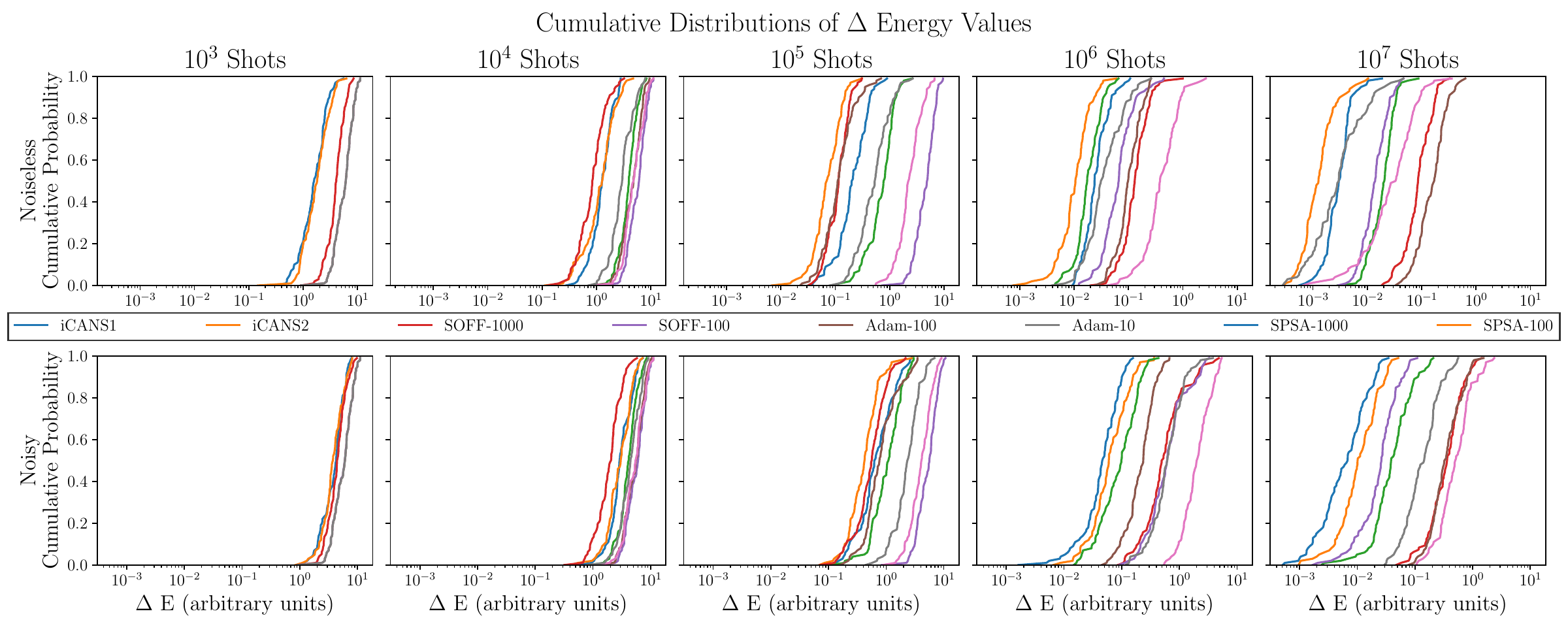}%
    \caption{Comparison of performance for the Heisenberg spin chain VQE task across one hundred random starts. Each panel shows the cumulative probability distribution of the energy difference from the ground state acheived by each optimizer for a different value of total shots $N$. Note that the further to the left a curve is, the better the optimizer has minimized the energy. }
    \label{fig:vqe_cumulative}
\end{figure*}

Here we show the cumulative distribution plots of the cost values or energies achieved by the optimizers we studied for the compilation task (Fig.~\ref{fig:compilation_cumulative}) and the Heisenberg spin chain VQE task (Fig.~\ref{fig:vqe_cumulative}) for various shot budgets. 



\end{document}